\documentclass[%
superscriptaddress,
preprint,
 amsmath,amssymb,
 aps,
 pra,
]{revtex4-1}

\usepackage[utf8]{inputenc}
\usepackage{graphicx} 
\usepackage{xcolor}
\usepackage{dcolumn}
\usepackage{bm}
\usepackage[symbol]{footmisc}

\begin{document}

\title{Full minimal coupling Maxwell-TDDFT: an ab initio framework for light-matter phenomena beyond the dipole approximation}

\author{Franco P. Bonafé\footnote[1]{Contact author: franco.bonafe@mpsd.mpg.de}}
\affiliation{Max Planck Institute for the Structure and Dynamics of Matter, Center for Free Electron Laser Science, Luruper Chaussee 149, 22761 Hamburg, Germany}
\author{Esra Ilke Albar}
\affiliation{Max Planck Institute for the Structure and Dynamics of Matter, Center for Free Electron Laser Science, Luruper Chaussee 149, 22761 Hamburg, Germany}
\author{Sebastian T. Ohlmann}
\affiliation{Max Planck Computing and Data Facility, Gießenbachstr. 2, 85748 Garching, Germany}
\author{Valeriia P. Kosheleva}
\affiliation{Max Planck Institute for the Structure and Dynamics of Matter, Center for Free Electron Laser Science, Luruper Chaussee 149, 22761 Hamburg, Germany}
\author{Carlos M. Bustamante}
\affiliation{Max Planck Institute for the Structure and Dynamics of Matter, Center for Free Electron Laser Science, Luruper Chaussee 149, 22761 Hamburg, Germany}
\author{Francesco Troisi}
\affiliation{Max Planck Institute for the Structure and Dynamics of Matter, Center for Free Electron Laser Science, Luruper Chaussee 149, 22761 Hamburg, Germany}
\author{Angel Rubio}
\affiliation{Max Planck Institute for the Structure and Dynamics of Matter, Center for Free Electron Laser Science, Luruper Chaussee 149, 22761 Hamburg, Germany}
\affiliation{Center for Computational Quantum Physics (CCQ), The Flatiron Institute, 162 Fifth Avenue, New York, New York 10010, USA}
\author{Heiko Appel\footnote[2]{Contact author: heiko.appel@mpsd.mpg.de}}
\affiliation{Max Planck Institute for the Structure and Dynamics of Matter, Center for Free Electron Laser Science, Luruper Chaussee 149, 22761 Hamburg, Germany}

\begin{abstract}
    We report the first \textit{ab initio}, non-relativistic QED method that couples light and matter self-consistently beyond the electric dipole approximation and without multipolar truncations. This method is based on an extension of the Maxwell-Pauli-Kohn-Sham approach to a full minimal coupling Hamiltonian, where the space- and time-dependent vector potential is coupled to the matter system, and its back-reaction to the radiated fields is generated by the full current density. The implementation in the open-source Octopus code is designed for massively-parallel multiscale simulations considering different grid spacings for the Maxwell and matter subsystems. Here, we show the first applications of this framework to simulate renormalized Cherenkov radiation of an electronic wavepacket, magnetooptical effects with non-chiral light in non-chiral molecular systems, and renormalized plasmonic modes in a nanoplasmonic dimer. We show that in some cases the beyond-dipole effects can not be captured by a multipolar expansion Hamiltonian in the length gauge. Finally, we discuss further opportunities enabled by the framework in the field of twisted light and orbital angular momentum, inelastic light scattering and strong field physics.
\end{abstract}

\maketitle

\section{Introduction}
Quantum electrodynamics (QED) is a fundamental field theory that describes all the phenomena associated with charged particles and photons, including well-known effects like the Lamb shift \cite{Lamb1947}, spontaneous emission \cite{Milonni1975,Milonni1984}, and the Casimir-Polder forces between metallic plates \cite{Casimir1948,buhmann2013dispersion}. QED effects have been particularly important in strong-field physics to describe phenomena such as electron-positron pair creation \cite{Schwinger1951} and vacuum birefringence \cite{Bialynicka-Birula1970,Karbstein2021}.
Just recently both experimental \cite{Thomas2016, Chikkaraddy2016, Orgiu2015} and theoretical evidence \cite{Fregoni2018, Schafer2019, Schafer2021} have suggested that strong
coupling between matter and modified vacuum in cavities leads to hybridized light-matter states. This development has paved the way for the emerging field of cavity quantum materials with relevant applications for a variety of quantum systems \cite{FriskKockum2019, Schlawin2022, Bloch2022}.
While the traditional QED approach is currently limited to relatively simple systems, such as few-electron atoms \cite{Glazov2019,Kosheleva2022,Zinenko2023,Kosheleva2020q}, including the quantum effects of both light and matter in a fully \textit{ab initio} framework becomes exponentially expensive unless approximated methods are used. 

To address these limitations, the theoretical framework of ab initio electronic structure theories for understanding light-matter interactions for realistic materials has evolved. Among these methods, time-dependent density-functional theory (TDDFT) \cite{Runge1984} is particularly successful due to its balance of accuracy and computational efficiency \cite{tddft2006}. In most of the cases treated within the TDDFT, an external electromagnetic field (such as a laser pulse) drives the system, which is usually coupled within the electric dipole approximation to the electrons. Quantum electrodynamical density functional theory (QEDFT) \cite{Ruggenthaler2014} has been developed to include the quantized electromagnetic fields as a photon-free QED functional \cite{Schafer2021}. However, semiclassical treatments that enable the coupling to light in arbitrary electromagnetic environments and recover some relevant QED phenomena are desirable.

In this sense, several semiclassical methods have been developed to account for these radiative effects in quantum systems, all of which demand solving the induced electromagnetic fields at some level, e.g. Maxwell-Bloch, Maxwell-Ehrenfest, and Maxwell-Ehrenfest+R methods \cite{Li2019}, multi-trajectory Ehrenfest \cite{Hoffmann2019,Hoffmann2019b}, coherent electric-electron dynamics \cite{Bustamante2021}, local radiation-reaction potentials \cite{Schafer2022}, Maxwell-Schrödinger \cite{Sukharev2023} and Maxwell-TDDFT \cite{Jestadt2019, Yamada2019}, Casida QEDFT \cite{Flick2019,Konecny2024}, and its combination with macroscopic QED among others.
Some methods, like the Ehrenfest+R or multitrajectory Ehrenfest approaches, include effects beyond the classical description, like spontaneous decay, but implementations in electron \textit{ab initio} calculations are still missing. In most of these methods, only the dipolar radiated field is calculated and its effect is coupled back into the Hamiltonian, which can account for radiative decay \cite{Bustamante2021, Schafer2022}, superradiance \cite{Bustamante2022} and coupling with cavities \cite{Hoffmann2019, Schafer2022}.

On the other hand, it has become clear that, in several areas of theoretical physics, light-matter interactions must be treated beyond the dipole approximation, with large spectroscopic applications, as this breaks dipole selection rules and creates opportunities for new driven light-matter systems \cite{Rivera2016}. Beyond-dipole effects have been demonstrated for X-ray and XUV spectroscopy \cite{Bernadotte2012, Lestrange2015, Foglia2022, List2015, Sakko2010, Aurbakken2024}, strong field phenomena \cite{Lindle1999, Forre2016, Iwasa2009, Forre2006, Forre2005, Jensen2020, KahvedZic2022, Lin2022}, magneto-optical effects \cite{Sun2019, Foglia2022}, coupling in chiral cavities \cite{Sun2022, Riso2023, Schafer2023}, molecules in nanoplasmonic cavities for enhanced spectroscopy \cite{Chen2019, Liu2019, Litman2023} and interactions with twisted light \cite{Picon2010, Quinteiro2017, Forbes2019c, Sun2022,Kosheleva2020}, just to name a few. Nonetheless, in these fields, a truncated multipolar expansion is the usual method of choice, which is origin-dependent  \cite{Lestrange2015} and its convergence with the multipole order can be slow \cite{Rivera2016}. Methods like the non-dipole strong field approximation Hamiltonian \cite{Jensen2020}, and non-truncated treatments for X-ray spectra calculations \cite{Foglia2022, Aurbakken2024, List2015} have gone beyond the multipolar expansion for these specific applications. A single tool that combines a general framework for beyond-dipole interactions in the time domain and can account for radiation-reaction has not been achieved so far.

Here, we report an open-source, parallel and efficient implementation of the real-space, real-time Maxwell-TDDFT method in full minimal coupling and showcase applications that highlight its use for QED, beyond-dipole spectroscopy, and both simultaneously. To the best of our knowledge, this is the first implementation of its kind to be reported. In particular, we show the renormalized Cherenkov emission from an electronic wavepacket by coupling back the emitted field into the quantum dynamics. In addition, we show the presence of magneto-optical effects driven by linearly polarized XUV light using benzene as an example, as an atomistic equivalent of dark transitions in nanorings. Finally, we demonstrate that a nanoparticle dimer coupled to the free-space electromagnetic environment has a plasmonic frequency shift as well as non-negligible phase shifts of the density and near field in the Fourier domain, by pure coupling with the transverse field induced by itself. To conclude, we present a few outlooks for the application of this tool to strong field phenomena, twisted light beams, and periodic systems.
\\
\indent

\section{Theory and Numerical Methods}

The SI system of units is used in the paper unless specified otherwise. As starting point, we consider the time-dependent Kohn-Sham equations using the full minimal coupling Hamiltonian in velocity gauge
\begin{align}
   \mathcal{H}^{\mathrm{f.m.c.}}=\frac{1}{2m}\left(-i\hbar \nabla + \frac{|e|}{c}\mathbf{A}(\mathbf{r},t)\right)^{2}+ \frac{|e|}{m} \mathbf{B}(\mathbf{r},t) \cdot \hat{\mathbf{s}} + V_{\mathrm{H}}[n](\mathbf{r},t)+ V_{\mathrm{xc}}[n](\mathbf{r},t)+V_{\text{nuc}}(\mathbf{r}, t).
   \label{eq:fullminham}
\end{align}
Here $\hbar$ is the reduced Planck constant, $m$ and $e$ are the mass and charge of the electrons, respectively, $c$ is the speed of light in vacuum, $\mathbf{A}(\mathbf{r},t)$ is the transverse vector potential and $\mathbf{B}(\mathbf{r},t)$ is the magnetic field associated with it, $\hat{\mathbf{s}}$ is the spin operator, $V_{\mathrm{H}}[n](\mathbf{r},t)$ and $V_{\mathrm{xc}}[n](\mathbf{r},t)$ are the Hartree and exchange-correlation potentials, respectively, $V_{\text{nuc}}(\mathbf{r}, t)$ is the Coulomb potential of the nuclei, and $n(\mathbf{r},t)$ is the electronic density~\cite{Ullrich}
\begin{equation}
n(\mathbf{r}, t)= \sum_{j=1}^{N_\mathrm{occ}}\sum_{\sigma=\pm \frac{1}{2}}\left|\varphi_{j}(\mathbf{r},\sigma,t)\right|^2,
\label{eq:density}
\end{equation}
where $\varphi_{j}(\mathbf{r},\sigma,t)$ are the Kohn-Sham wavefunctions with index $j$ enumerating the $N_\mathrm{occ}$ occupied orbitals and $\sigma$ being spin. The transverse vector potential $\mathbf{A}(\mathbf{r},t)$ can have both external and matter contributions
\begin{align}
    \mathbf{A}(\mathbf{r},t)=\mathbf{A}_{\mathrm{mat}}(\mathbf{r},t)+\mathbf{A}_{\mathrm{ext}}(\mathbf{r},t).
   \label{eq:vecpotsum}
\end{align}
The matter-induced vector potential $\mathbf{A}_{\mathrm{mat}}(\mathbf{r},t)$ is a transverse field that depends on the properties of matter. We note that in most of the cases $\mathbf{A}_{\mathrm{mat}}(\mathbf{r},t)$ does not have an analytical expression and needs to be calculated numerically by solving Maxwell's equations self-consistently along with the electronic dynamics. 
Following Ref.~\cite{Jestadt2019} we choose the Riemann-Silberstein representation to propagate the electromagnetic fields, where the Riemann-Silberstein vector $\mathbf{F}(\mathbf{r},t)$ is defined as
\begin{align}
\mathbf{F}^\pm(\mathbf{r},t)= \sqrt{ \epsilon_0/2}\left(\mathbf{E}(\mathbf{r},t) \pm i c \mathbf{B}(\mathbf{r},t)\right),
\label{eq:rsvector}
\end{align}
where $\epsilon_0$ is the vacuum permittivity. For the remainder of this work we use only the positive definition: $\mathbf{F} := \mathbf{F}^+$.
According to this definition, the Faraday and Ampère laws are unified into a single equation of motion
\begin{align}
i \partial_t \mathbf{F}(\mathbf{r}, t)= c \boldsymbol{\nabla} \times \mathbf{F}(\mathbf{r}, t)-\frac{i}{\sqrt{2 \epsilon_0}} \mathbf{J}(\mathbf{r}, t),
\label{eq:rsmaxwelleqs}
\end{align}
while the two Gau{\ss} laws are similarly combined
\begin{align}
 \nabla \cdot \mathbf{F}(\mathbf{r}, t) = \sqrt{\frac{1}{2 \epsilon_0}} n(\mathbf{r},t).
 \label{eq:rsgausseqs}
\end{align}
The source term $\mathbf{J}(\mathbf{r},t)$ in \eqref{eq:rsmaxwelleqs} is the total current density given by \cite{Berestetskii1971,Jestadt2019}
\begin{align}
\nonumber
\mathbf{J}(\mathbf{r},t)&=\frac{e\hbar }{2\mathrm{i}m}\sum_{j=1}^{N_\mathrm{occ}}\sum_{\sigma=\pm \frac{1}{2}} \left( \left[\boldsymbol{\nabla}\varphi^{\dagger}_j(\mathbf{r},\sigma,t)\right]\varphi_j(\mathbf{r},\sigma,t) - \varphi_j^{\dagger}(\mathbf{r},\sigma,t)\left[\boldsymbol{\nabla}\varphi_j(\mathbf{r},\sigma,t)\right]\right)
-\frac{e^2}{m c} \mathbf{A}(\mathbf{r},t) n(\mathbf{r},t) \\
&- \frac{e\hbar}{2m} \boldsymbol{\nabla} \times\sum_{j=1}^{N_\mathrm{occ}}\sum_{\sigma=\pm \frac{1}{2}}\varphi_j^{\dagger}(\mathbf{r},\sigma,t)\hat{\mathbf{s}}\varphi_j(\mathbf{r},\sigma,t),
\label{eq:currentfull}
\end{align}
where the first, second, and third terms are the paramagnetic, diamagnetic, and magnetization current densities, respectively.

The vector potential can now be computed from the time-propagated magnetic field $\mathbf{B}(\mathbf{r},t) = \sqrt{2\mu_0}\Im(\mathbf{F}(\mathbf{r},t))$ by solving the Poisson equation for $ \nabla \times \mathbf{A}(\mathbf{r},t) = \mathbf{B}(\mathbf{r},t)$
\begin{align}
\mathbf{A}(\mathbf{r},t)=\frac{1}{4 \pi} \int_V \frac{\boldsymbol{\nabla}^{\prime} \times \mathbf{B}\left(\mathbf{r}^{\prime},t\right)}{\left|\mathbf{r}-\mathbf{r}^{\prime}\right|} \mathrm{d}^3 r^{\prime}-\frac{1}{4 \pi}\oint_S \hat{\mathbf{n}}^{\prime} \times \frac{\mathbf{B}\left(\mathbf{r}^{\prime},t\right)}{\left|\mathbf{r}-\mathbf{r}^{\prime}\right|} \mathrm{d}^2 r^{\prime}.
\label{eq:vecpotmag}
\end{align}
The first term can be efficiently calculated using a Poisson solver, as it corresponds to the solution of the Poisson equation for a simulation box with volume $V$. The second integral in Eq. \eqref{eq:vecpotmag} is a surface integral, where $\hat{\mathbf{n}}^{\prime}$ denotes the normal to the surface $S$ of the simulation box. This term becomes necessary if $\mathbf{B}\left(\mathbf{r}^{\prime},t\right)$ does not vanish at the surface of the simulation box. In Coulomb gauge, however, eq. \eqref{eq:vecpotmag} can be written as a function of the instantaneous magnetic field \cite{Stewart2003}
\begin{align}
    \mathbf{A}\left(\mathbf{r},t\right) = \nabla \times \frac{1}{4\pi} \int \frac{\mathbf{B}\left(\mathbf{r}',t\right)}{|\mathbf{r} - \mathbf{r}'|} \,\mathrm{d}^3 r^{\prime},
    \label{eq:vecpotmagcoulomb}
\end{align}
which is much more convenient for the numerical implementation (see Supporting Information).
\\
\indent
As only the transverse component of the electromagnetic fields is coupled back into the matter Hamiltonian, the longitudinal fields do not need to be calculated as a contribution from the Maxwell propagation, as long as the full matter system is interacting via the Coulomb potential. Hence, we do not account for the initial charge density in the Maxwell simulations, instead, we set the initial field to zero and let it evolve in time. In this way, the Gau{\ss} law is fulfilled with the charge density \textit{difference} with respect to the ground state
\begin{align}
\nabla \cdot \mathbf{E} (\mathbf{r},t')=\frac{n(\mathbf{r},t')-n(\mathbf{r},t=0)}{\varepsilon_0}.
\label{eq:gausselecfield}
\end{align}
Therefore, the Gau\ss~ law is automatically fulfilled and can be used as a side condition to test the method's accuracy (\textit{vide infra}). Another side condition that needs to be fulfilled automatically is the continuity equation: $ \partial_t \rho (\mathbf{r},t) + \boldsymbol{\nabla} \cdot \mathbf{J}(\mathbf{r},t)=0 $ (where $\rho = -e n$).
\\
\indent
The full minimal coupling Hamiltonian \eqref{eq:fullminham} gives access to light-matter interactions up to all multipolar orders.
However, a Taylor expansion of the vector potential and magnetic field in Eq. \eqref{eq:fullminham} around the center $\mathbf{r}_0$ can be performed, which leads to the well-known multipolar-expanded Hamiltonian.
This approach allows us to distinguish the effects arising from different orders and types of electromagnetic fields.
We have then implemented the multipolar Hamiltonian in length gauge up to the second order, which reads as follows \cite{Jestadt2019}
\begin{align}
\nonumber
\mathcal{H}^{\mathrm{m.e.}} &= -\frac{\hbar^2}{2m} \nabla^2 + |e|  \mathbf{r} \cdot \mathbf{E}_{\perp}\left(\mathbf{r}_{0},t\right) + \mathrm{i} \frac{|e|}{2 m} \mathbf{B}\left(\mathbf{r}_{0},t\right) \cdot\left(\mathbf{r} \times \boldsymbol{\nabla}\right) + \left.\frac{1}{2} |e|(\mathbf{r} \cdot \boldsymbol{\nabla}) \mathbf{r} \cdot\left(\mathbf{E}_{\perp}(\mathbf{r},t)\right)\right|_{\mathbf{r}=\mathbf{r}_{0}}+ \\
&+V_{\mathrm{H}}[n](\mathbf{r},t)+ V_{\mathrm{xc}}[n](\mathbf{r},t)+V_{\text{nuc}}(\mathbf{r}, t),
\label{eq:multipolarham}
\end{align}
which in addition to the commonly used electric dipole term $|e|  \mathbf{r} \cdot \mathbf{E}\left(\mathbf{r}_{0},t\right)$, also has a magnetic dipole contribution,
\begin{align}
    \mathrm{i} \frac{|e|}{2 m}\mathbf{B}\left(\mathbf{r}_{0},t\right) \cdot(\mathbf{r} \times \boldsymbol{\nabla})  = - \frac{|e|}{2 m} \mathbf{B}\left(\mathbf{r}_0,t\right) \cdot \mathbf{\hat{L}},
    \label{eq:magdip}
\end{align}
and the electric quadrupole term,
\begin{align}
\left.\frac{1}{2} |e|(\mathbf{r} \cdot \boldsymbol{\nabla}) \mathbf{r} \cdot\left(\mathbf{E}_{\perp}(\mathbf{r},t)\right)\right|_{\mathbf{r}=\mathbf{r}_{0}}  = \frac{1}{2} |e| \sum_{ij}r_i \mathbb{Q}_{ij}  r_j .
\label{eq:elecquad}
\end{align}
Here $\mathbf{L}$ is orbital angular momentum operator and $\mathbb{Q} _{i j}=\left.\partial_i \mathbf{E}_{\perp,j}\left(\mathbf{r},t\right)\right|_{\mathbf{r}=\mathbf{r}_0}$ is the electric field gradient tensor.

In summary, the initial conditions for the Kohn-Sham orbitals $\varphi^0$ and Riemann-Silberstein vectors $\mathbf{F}^0$ are given by
\begin{align}
    \left( -\frac{\hbar^2}{2m}\nabla^2 + V_{\mathrm{H}}[n](\mathbf{r}) + V_{\mathrm{xc}}[n](\mathbf{r}) + V_{\text{nuc}}(\mathbf{r}) \right) & \varphi_j^0(\mathbf{r},\sigma) = E_j \varphi_j^0(\mathbf{r},\sigma), \\
    \mathbf{F}^0(\mathbf{r}) &= 0
\end{align}
and their coupled dynamics are defined by the following equations, and summarized in Fig. \ref{fig:grids_coupling}
\begin{align}
    i\hbar \partial_t \varphi_j(\mathbf{r},\sigma,t) &= \mathcal{H}^{\mathrm{f.m.c.}} \varphi_j(\mathbf{r},\sigma,t), \\
    i \partial_t \mathbf{F}(\mathbf{r}, t) &= c \boldsymbol{\nabla} \times \mathbf{F}(\mathbf{r}, t)-\frac{i}{\sqrt{2 \epsilon_0}} \mathbf{J}(\mathbf{r}, t).
\end{align}
\subsection{Numerical Methods}

We solve the coupled Kohn-Sham-Maxwell equations as given by the full minimal coupling Hamiltonian within TDDFT as implemented in the open-source code Octopus \cite{Tancogne-Dejean2020}. The Kohn-Sham equations are solved in real space and real time. The Maxwell equations in Riemann-Silberstein form are discretized in space using the same finite differences as for the Kohn-Sham equations and in time using an exponential propagation scheme \cite{Jestadt2019}. Both the Kohn-Sham equations and the Maxwell equations have been implemented as different systems coupled via the new multi-system framework in Octopus \cite{Tancogne-Dejean2020}. The use of the new framework has not only enabled the interaction among more than two systems, giving more freedom for the types of simulation, but also has opened the way to couple several electronic and Maxwell systems and also to extend this to other levels of theory such as, e.g., continuum electromagnetic models, and density-functional tight-binding calculations, having potential for truly multiscale simulations.

The current implementation allows the inclusion of different boundary conditions for solving Maxwell's equations: zero fields, constant fields, and absorbing boundary conditions using a perfectly matched layer (PML). Moreover, plane waves can be introduced into the simulation box by using Dirichlet boundary conditions. For this paper, we have used PML absorbing boundaries for the Maxwell simulations, to simulate emission of the scattered fields in free space. This, however, demands the use of a much larger Maxwell box than the matter box, so that the PML and the matter system do not overlap in space.

\begin{figure}
    \centering
    \includegraphics[scale=0.8]{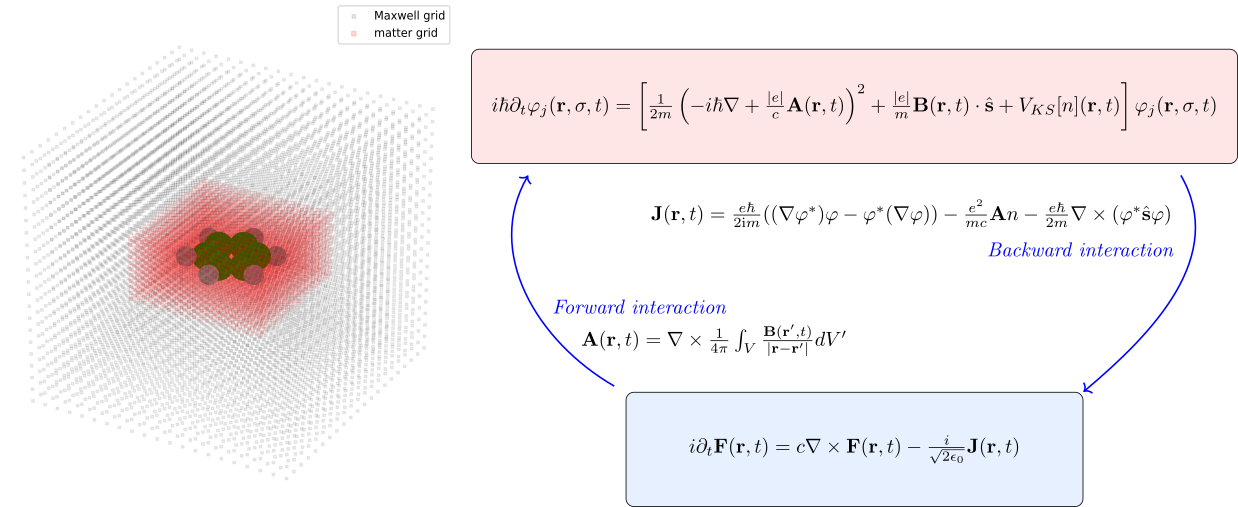}
    \caption{Sketches of the Maxwell and matter simulation boxes (left) and the forward-backward self-consistent coupling scheme (right). For this sketch, each grid point that is plotted represents a cube of 8$\times$8$\times$8 grid points in the real calculation.}
    \label{fig:grids_coupling}
\end{figure}

The coupling between systems supports different grids with arbitrary shapes and spacings (see Figure \ref{fig:grids_coupling}). The transfer of quantities (called regridding) from one grid to the other is a central operation in our framework, as it is implemented using both  linear and nearest-neighbor interpolations on the overlapping parts of the grids. It also supports grids with different parallel distributions. With this approach, we can adapt the grid spacing and shape for the different systems, thus saving computational effort.

Finally, the coupling of systems supports various time steps through the multi-system framework, where the mapped quantities are interpolated in time to ensure that the order of the time propagation methods is retained. Mostly, a 4th-order Taylor expansion of the exponential operator (and thus also a 4th-order time interpolation) is used. This feature allows us to efficiently capture different temporal scales of the different systems (as the Maxwell equations have naturally a faster timescale and demand a smaller timestep than the electron dynamics).

\subsection{Validation of the Maxwell solver}
In this subsection, we illustrate the validation of our Maxwell solver. We first consider in \ref{sec:rad_oscillator} the radiation of an electron in the presence of a harmonic potential. We show that the numerical simulation is in perfect agreement with analytical results. As a next step, we verify in \ref{sec:helmholtz} the numerical solution of the Poisson equation \eqref{eq:vecpotmag} and check the accuracy of the Helmholtz decomposition. In \ref{sec:benzene_origin}, we investigate the effects of origin dependence on the dipole moment of benzene driven by a laser field. And finally, in \ref{sec:pseudo_vs_all} we discuss how the observables calculated with the usage of the non-local potential (widely used pseudopotentials) deviate from the case of local (all-electron) potentials.

\subsubsection{TDDFT to Maxwell: Radiation from the harmonic oscillator}
\label{sec:rad_oscillator}

\begin{figure}
    \centering
    \includegraphics[scale=0.5]{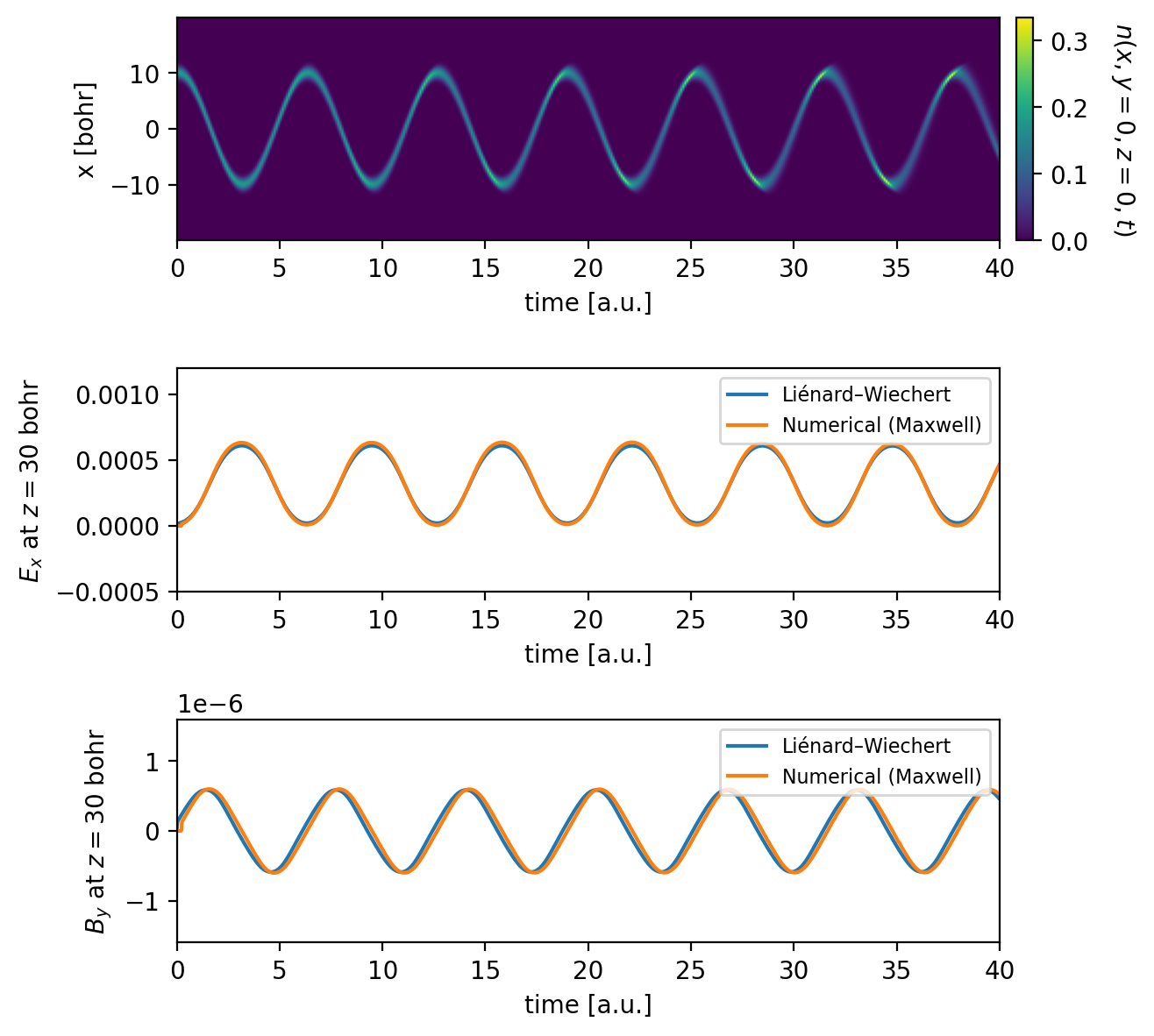}
    \caption{Dynamics of a wavepacket in a harmonic oscillator (top) and its generated electromagnetic fields as a function of time (bottom). The top panel depicts a 1D slice of the density in the $x$ axis ($y=0, z=0$) as a function of time. From the full volumetric density the dipole moment is calculated and is fed into the Liénard–Wiechert equations for the analytical fields, fulfilling the role of the particle position $\mathbf{r}_p$. The middle panel shows the $x$ component of the time-propagated electric field and its analytical counterpart at the (0,0,30 bohr) point as a function of time, while the magnetic field $y$ component is compared for both cases in the bottom panel.}
    \label{fig:ho_mxll}
\end{figure}

As a test of the implementation, we have studied the electromagnetic fields induced by an electron wavepacket oscillating in a harmonic potential in three dimensions. Namely, we consider an initial wavefunction as the solution in the displaced quadratic potential $V = \frac{1}{2}((x-10)^2 + y^2 + z^2)$, and for the dynamics we displace the potential from the origin, triggering the electronic dynamics. The wavepacket is then propagated for 40 a.u. of time, while the spatially-resolved radiation from the wavepacket is obtained by solving Maxwell's equations with absorbing boundary conditions (the details of the grids used can be found in the Supporting Information). The results of the simulation are depicted in Fig.~\ref{fig:ho_mxll}.
 
Far from the source wavepacket, where the effect of the Gaussian charge distribution is smaller, the generated electromagnetic fields have to match the Liénard–Wiechert (LW) fields of a point-charged particle. The electric and magnetic fields described by the LW formulas are given by \cite{griffiths2023introduction}:
\begin{align}
        \mathbf{E}(\mathbf{r}, t) = \frac{q}{4 \pi \epsilon_0}\frac{\tilde{r}}{(\tilde{\mathbf{r}} \cdot \mathbf{u})^3} \left[ (c^2 - v^2) \mathbf{u} + \tilde{\mathbf{r}} \times (\mathbf{u} \times \mathbf{a}) \right] \; \; ; \; \; 
        \mathbf{B}(\mathbf{r}, t) = \frac{1}{c} \frac{\tilde{\mathbf{r}}}{\tilde{r}} \times \mathbf{E}(\mathbf{r}, t).
\label{eq:lwfield}
\end{align}
Here, $\mathbf{r}$ represents the position of the field, $\tilde{\mathbf{r}} = \mathbf{r} - \mathbf{r}_p$, where $\mathbf{r}_p$ is the source position, considered in our calculations as the mean value position of the wave packet. The vector $\mathbf{u} = c \tilde{\mathbf{r}} / \tilde{r} - \mathbf{v}$, where $\mathbf{v}$ is the velocity vector of the source and $\mathbf{a}$ is its acceleration vector.

In the bottom panels of Fig.~\ref{fig:ho_mxll} the simulated and the LW fields are compared at one spatial point relatively far from the source, showing almost perfect agreement, the small deviation arising from the finite width of the wavepacket. As the results have not been rescaled by any factor, this test shows that the Maxwell time propagation with an electronic system as the source is correct and consistent with the employed unit system in our code.

\subsubsection{Maxwell to TDDFT: magnetic vector potential and Helmholtz decomposition}
\label{sec:helmholtz}

To verify that the vector potential from a Maxwell system is properly coupled to the electronic system, we have propagated a plane wave pulse in a Maxwell box, imposing the initial values at the boundaries (Dirichlet conditions), and calculated its vector potential using Eq. \eqref{eq:vecpotmag}, which implies solving a Poisson equation for the magnetic field. The comparison of the calculated $\mathbf{A}(\mathbf{r},t)$ and the analytical expression from the plane-wave ($\mathbf{A}=-\partial_t \mathbf{E}$) proves that the calculation is valid in most of the box (see Supporting Information), especially for points which are far from the boundaries, where the error is larger due to the approximate description of the derivatives at the boundaries (see description of the methods for the Helmholtz decomposition in the Supporting Information).

We also verified that transverse and longitudinal electric field components are correct. These components were calculated using the Helmholtz decomposition for an oscillating point dipole, described as a space-dependent spatial current, where the emitted fields are absorbed by the PML absorbing boundaries. The accuracy of the decomposition for different box sizes and PML widths can be evaluated by comparing the total field with the sum of the transverse and longitudinal fields (see Supporting Information). This comparison is useful to establish the minimum box size and maximum wavelength that can be simulated within a certain tolerance, for a charge distribution of a given spatial extent.

\subsubsection{Origin-independence}
\label{sec:benzene_origin}

\begin{figure}
    \centering
    \includegraphics[scale=0.4]{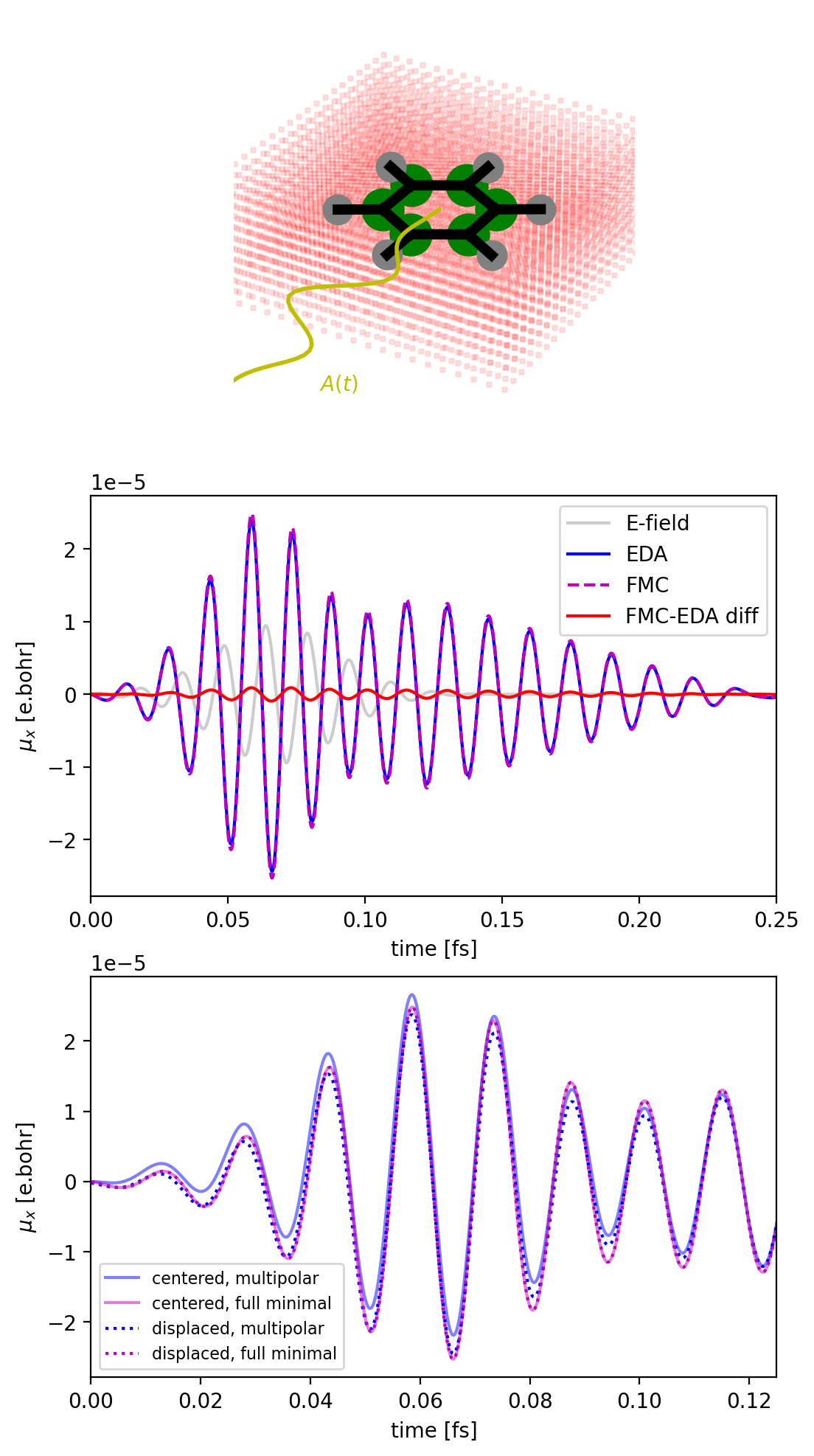}
    \caption{Laser-driven benzene molecule by a pulse with central frequency $\omega=270$ eV, comparing full minimal coupling with the truncated multipolar coupling at the magnetic dipole/electric quadrupole level. Top panel: sketch of the simulation box and incoming field. Middle panel: comparison of the time-dependent dipole moment, $x$ component, in electric dipole, full minimal coupling and its difference, for the benzene molecule centered at the (0,0,0) point (the external field time-trace is included for reference). Bottom panel: comparison of the dipole moment dynamics in full minimal coupling and with the multipolar expansion including electric dipole, magnetic dipole and electric quadrupole terms, with the molecule placed at two different locations: in one case (centered) with the center of mass at $\mathrm{r}_0=(0,0,0)$ and in the other (displaced) with $\mathrm{r}_0=(8,8,0)$ bohr.}
    \label{fig:benzene_extsource}
\end{figure}

It is well known that truncating the light-matter interaction at a given multipole order beyond dipole leads to origin-dependent observables \cite{Lestrange2015}. As the multipolar expansion is a common choice in electronic structure calculations, it is important to test its accuracy. Here we compare different coupling levels for the time-propagation of a benzene molecule located at two different positions in the box, driven by an XUV pulse of frequency 270 eV which propagates along the $y$ axis and is depicted in Fig. \ref{fig:benzene_extsource}. The two locations for the center of mass $\mathbf{r}_0$ (see eq. \eqref{eq:multipolarham}) are namely the origin (0, 0, 0) (referred to as \textit{centered}) and at (8, 8, 0) bohr (\textit{displaced}). First of all, we observe non-negligible differences between the full minimal coupling and the dipolar level, shown in the top panel, which will be analyzed in detail later. 

Here we focus on the comparison between full minimal coupling and a multipolar expansion that considers electric dipole, magnetic dipole, and electric quadrupole couplings. In all these cases we consider only forward-coupling of the plane wave with the electronic system, ignoring the back reaction. In the bottom panel of Fig. \ref{fig:benzene_extsource}, the dipole moment of the centered and displaced molecules are compared. As the plane wave propagates along the $y$ direction, one molecule experiences the effect of the external field earlier than the other. To correct for this delay and to have comparable time traces, the dipole moment of the displaced system has been translated in the time axis by an offset of $t_{\mathrm{off}} = (8 \; \mathrm{bohr}) / c = 0.058$ a.u. It is visible that the results in full minimal coupling match perfectly, while the simulations with truncated multipolar treatment show deviations. Considering that the beyond-dipole effects induced by a plane wave of this frequency are still within a few percent, the expected effects when the multipolar truncation is done on much more inhomogeneous fields (like a plasmonic near field) are expected to be much more important, potentially leading to incorrect conclusions.

\subsubsection{Pseudopotentials vs. all-electron}
\label{sec:pseudo_vs_all}

\begin{figure}
    \centering
    \includegraphics[scale=0.5]{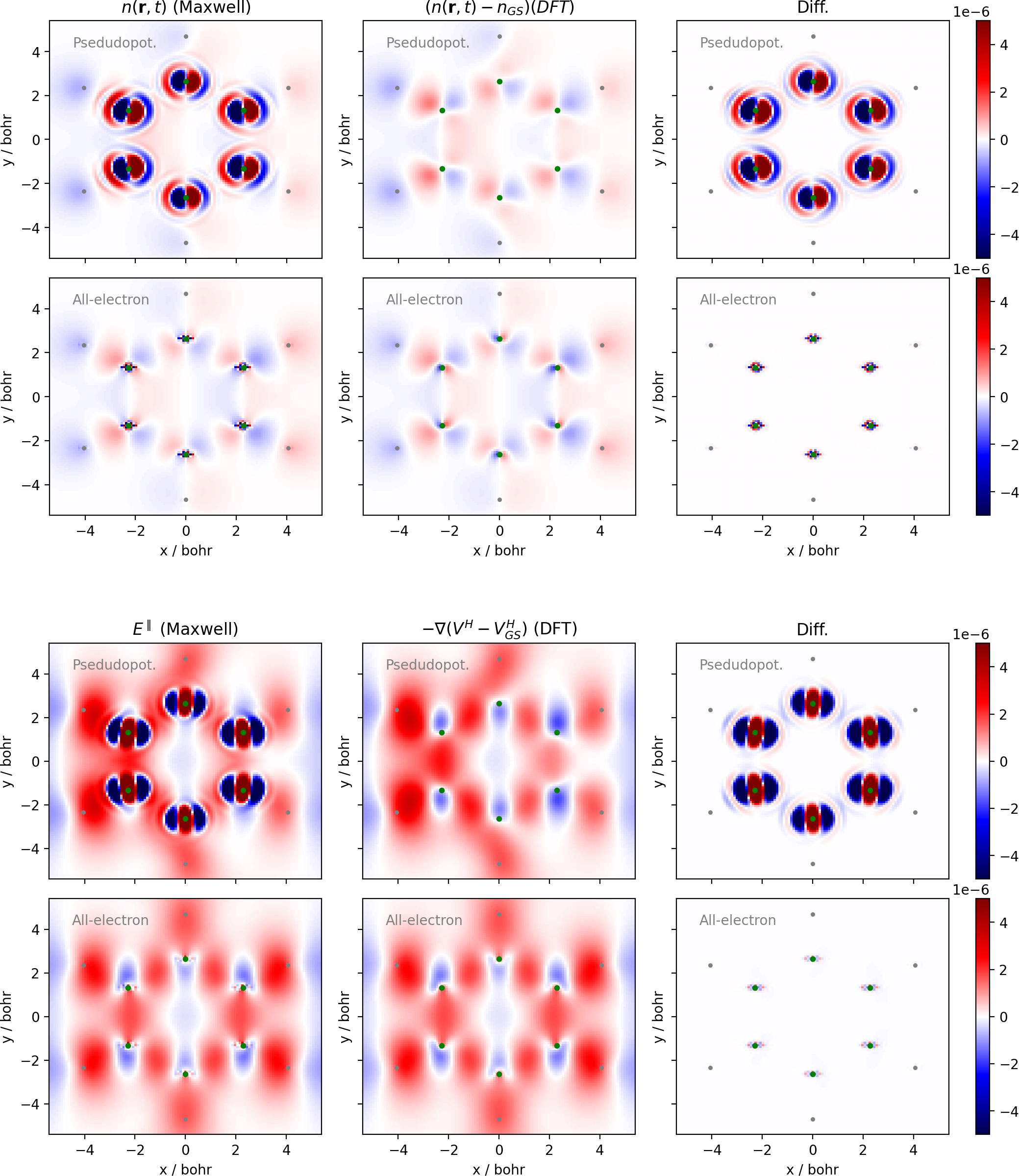}
    \caption{Comparison of slices in the $xy$ plane of the electron density (top panels) and longitudinal electric field (bottom panels) as calculated from the DFT subsystem and from the electromagnetic fields in the Maxwell solver, when using pseudopotentials and for an all-electron calculation. In all cases, the simulation consisted of a benzene molecule centered at (0,0,0) driven in electric-dipole-approximation by an external pulse of frequency $\omega = 7.2$ eV, tuned to the HOMO-LUMO transition. The snapshots are taken at time $t=12$ a.u.}
    \label{fig:benzene_pseudo}
\end{figure}

The path ambiguity in the gauge transformation of the non-local part of the atomic pseudopotentials is well known, and possible solutions have been proposed, although with limited applicability in realistic scenarios \cite{Ismail-Beigi2001,Pickard2003}. In specific cases, e.g. when a kick is applied, and therefore a vector potential is not present in the simulation, the problem can be overcome by transforming the momentum operator \cite{Varsano2009}. However, in general, there is not an efficient method to overcome this limitation, and in this work, we focus on calculations that do not involve a non-local potential, by using an all-electron description. To understand the error produced by the use of pseudopotentials, we compare simulations using both descriptions for a benzene molecule illuminated with a 7-eV external pulse, while solving Maxwell's equations. In Fig.~\ref{fig:benzene_pseudo} we compare two quantities that can be computed from the electronic and the Maxwell systems, to see if they are consistent under the two electronic descriptions. We namely compare the density and calculate both as defined in Eq.~\eqref{eq:density}, as well as by solving Eq.~\eqref{eq:gausselecfield} (top panels); and the longitudinal electric field, which can be computed from the Helmholtz decomposition of the total electric field, as well as from the gradient of the time-dependent Hartree potential in DFT: $ \mathbf{E}^{\parallel} = -\boldsymbol{\nabla}(V^H - V^H_{GS}) $. We see that while the all-electron quantities agree when calculated from the different physical systems (albeit small deviations close to the atomic nuclei arising due to the grid discretization), the same quantities differ inside the pseudopotential region for the calculations using pseudopotentials. Due to the computational cost of all-electron calculations, in future works it would be desirable to find a solution for the gauge invariance when pseudopotentials are used. For that purpose, the visual representation of the differences presented here could be a hint to find an adequate approach to overcome the path ambiguity.

\section{Cherenkov radiation from an electronic wavepacket}

The Vavilov-Cherenkov radiation is the emitted radiation from a particle that is traveling faster than the speed of light in a medium \cite{Cherenkov1937, Frank1991}. 
As the radiated Cherenkov field is highly inhomogeneous and is emitted in the vicinity of the particle, backward-coupling of the induced radiation in the dynamics of the particle itself could, in principle, exist. As probing the Cherenkov angle is crucial to extract information from the superluminal travel of particles in a wide range of applications \cite{Bolotovskii2009}, beyond-dipole self-consistent corrections to the spatial distribution of the wavepacket as well as its dynamics would have practical applications.

To test this hypothesis, we perform simulations of an electronic wavepacket traveling at a higher velocity than the phase velocity of light in a medium.
From these simulations we analyze the emitted radiation with and without back-reaction of such radiation on the wavepacket dynamics. The initial condition of the Gaussian electronic wavepacket is the following:
\begin{equation}\label{cherenkov_initial_cond}
   \varphi(x,y,z,t=0) = C_0  e^{-i p (x - x_0)} e^{-\frac{1}{4}\left(
    \frac{x^2}{(\Delta x)^{2}} + \frac{y^{2}}{(\Delta y)^{2}} + \frac{z^{2}}{(\Delta z)^{2}} \right)},
\end{equation}
where $C_0$ denotes the amplitude while $\Delta x$, $\Delta y$ and $\Delta z$ indicates the width of the wavepacket in three directions. The term $e^{-i p (x- x_0)}$ shows that the wavepacket has momentum $p$ along the $x$ direction, initially centered at $x_0$. In our simulations, $p=2.5$ a.u., while the medium has a refractive index of 100, and consequently the phase velocity of light is $v_l = 1.37$ a.u.

To observe the Cherenkov effect, the wavepacket should not disperse too quickly during the dynamics. This implies keeping the momentum spread $\delta p$ low enough without making the position spread $\delta x$ too large to still fit in the simulation box. The momentum itself sets a constraint in the spacing used, due to the spatial modulations of the wavefunction. After choosing the parameters carefully, we employed a simulation box of size 60$\times$18$\times$18 bohr$^3$ with a uniform spacing of 0.3 bohr. The wavepacket width is $\Delta x = 5.5$ bohr in the $x$ direction, and $\Delta y = \Delta z = 2.5$ bohr in the $y$ and $z$ directions. The wavepacket moves in the positive $x$ direction. We offset the starting point of the particle by $-30$ nm in the $x$ direction.

\begin{figure}
    \centering
    \includegraphics[scale=0.5]{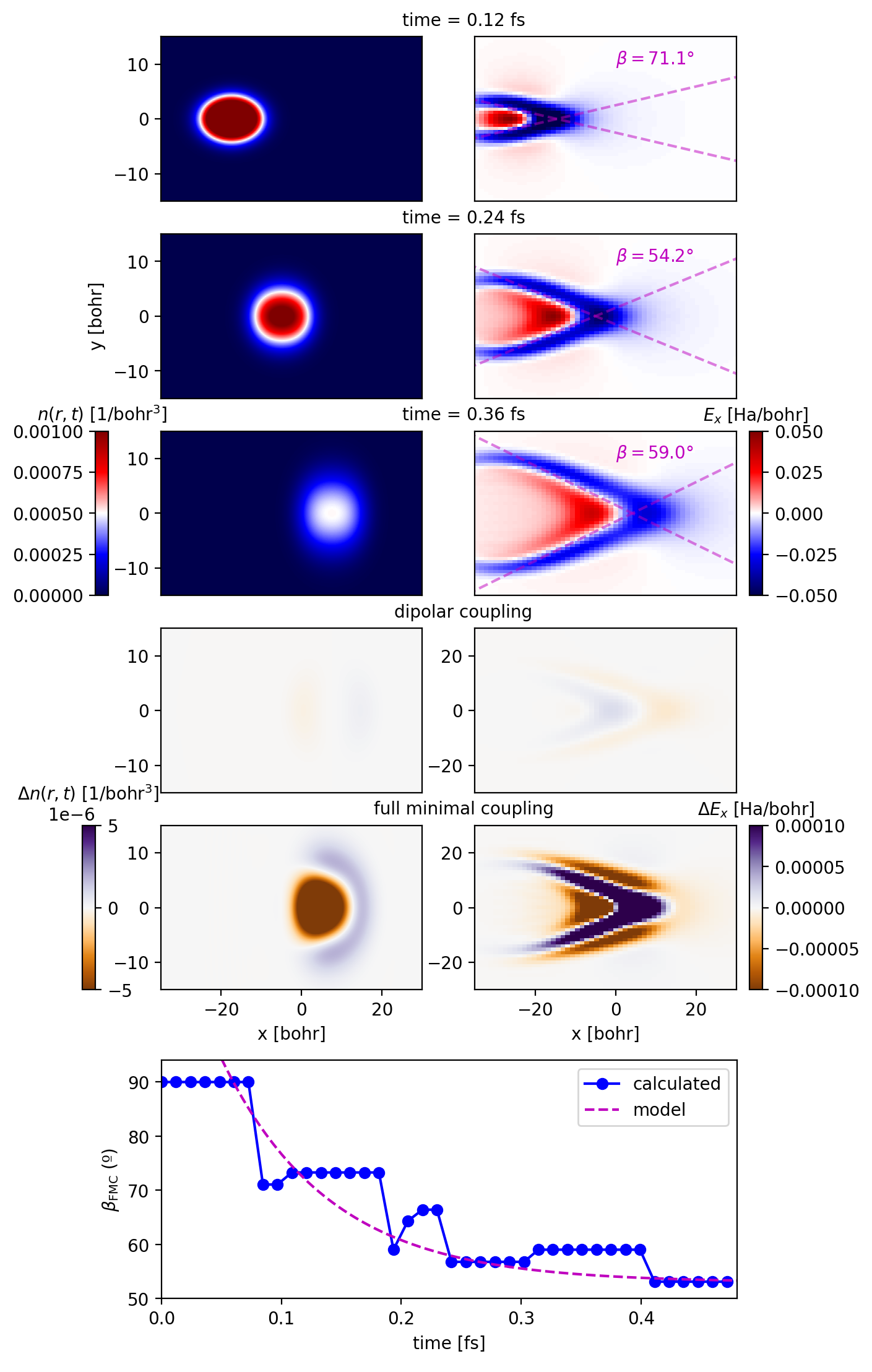}
    \caption{Cherenkov radiation: electron density and transverse electric field for different timesteps (top); corrections that arise in dipolar and full minimal coupling (middle); and dynamics of the Cherenkov angle formation (bottom). For the three time snapshots chosen in the top panels, the wavepacket is propagating freely without any external drive and without back-reaction. It can be seen how the wavepacket disperses while the radiation forms the well-known cone. The corrections are plotted with the same colorscale, which shows a minor effect from the dipolar coupling, which also shows variations along the axis of propagation of the wavepacket, while a rich angular-dependence arises in our full minimal coupling approach. The angle dynamics (bottom) show the time evolution of the angle measured from the simulations (calculated) together with an exponential fit (model) with a characteristic time of 0.09 fs.}
    \label{fig:cherenkov_plot}
\end{figure}

As it can be seen in Fig.~\ref{fig:cherenkov_plot}, as time evolves the wavepacket disperses, while the electric field ($x$-component, in this case) starts to develop the well-known Cherenkov cone. The non-dispersive Cherenkov angle is given by $\beta_{non-disp} = \arctan(v_l/p) = 56.8^{\circ}$. Here we study the birth of the Cherenkov radiation and the time evolution of the angle, being first close to $90^{\circ}$ (initial field aligned with the axis of propagation), as it develops and tends to $\approx 54^{\circ}$. The slight difference with the non-dispersive value can be attributed to the dispersion of the wavepacket and the numerical resolution of the grid.

Then, we included the back-action corrections, which are also shown in Fig.~\ref{fig:cherenkov_plot} both for dipolar and full minimal couplings. As can be seen, the corrections to the density and the emitted field are around 0.1\% with dipolar back-action (spatial average of the radiated field), but over 3\% with full minimal coupling back-action. The dipolar correction shows the expected dipolar pattern in the direction of propagation of the wavepacket, namely a slight shift in space of the density (which can be interpreted as a slightly larger velocity with respect to the non-coupled case). However, the correction with full minimal coupling shows a rich spatial distribution, creating a slight accumulation on the front and lateral sides, and a strong depletion on the back of the wavepacket, thereby changing its shape. The emitted field also shows an increase of the wavefront width and a decrease (in absolute values) of the field at the tip of the cone. In the literature, the corrections to Cherenkov dynamics have been addressed in a non-perturbative QED approach \cite{Roques-Carmes2018}. The present methodology has the potential to simulate the corrected non-perturbative dynamics of arbitrary particles described at the first principles level and to study the angular distribution of radiation for relevant scenarios.

\section{Magnetooptical response of non-chiral molecules with non-chiral soft X-rays}

It is known that nondipole corrections in X-ray spectroscopy, particularly in X-ray absorption (XAS) are around 5-10\% for dipole-allowed transitions in the soft X-ray region \cite{List2015}. Numerous works have proposed methods to calculate the nondipole response function from first principles without truncation of the multipolar expansion (e.g. in full minimal coupling) \cite{List2015, Foglia2022, Aurbakken2024}. Here, we analyze the beyond-dipole effects in real-space and real-time, by simulating the soft X-ray excitation of a single benzene molecule considering different angles of incidence, as defined by the wavevector. In this section we omit the back-reaction as it is very small (only a frequency shift of around 0.05\% is observed), therefore the electromagnetic field used is only of external origin.

The setup is the same as discussed in section \ref{sec:benzene_origin}. By considering an incident beam with both propagation and polarization vectors in the plane of the molecule ($\mathbf{k} = k \mathbf{e}_y$, $\mathbf{A} = A(r,t) \mathbf{e}_x$), the atomic dipole oscillations are dephased along the propagation direction, which is not visible in the dipolar approximation. This phenomenon is not observed at normal incidence, i.e. when the propagation of the beam is along the direction perpendicular to the molecule ($\mathbf{k} = k \mathbf{e}_z$).

This effect can be understood as the atomic-scale version of the phase retardation effects on the optical spectra of nanorings \cite{Hao2008}. As in the case of the nanorings, the in-plane beam excites higher-order modes which would otherwise be ``dark'' or dipole-forbidden. As the first order beyond the electric dipole is the electric quadrupole and the magnetic dipole, we examine the dynamics of the induced orbital angular momentum as follows.

\begin{figure}
    \centering
    \includegraphics[scale=0.6]{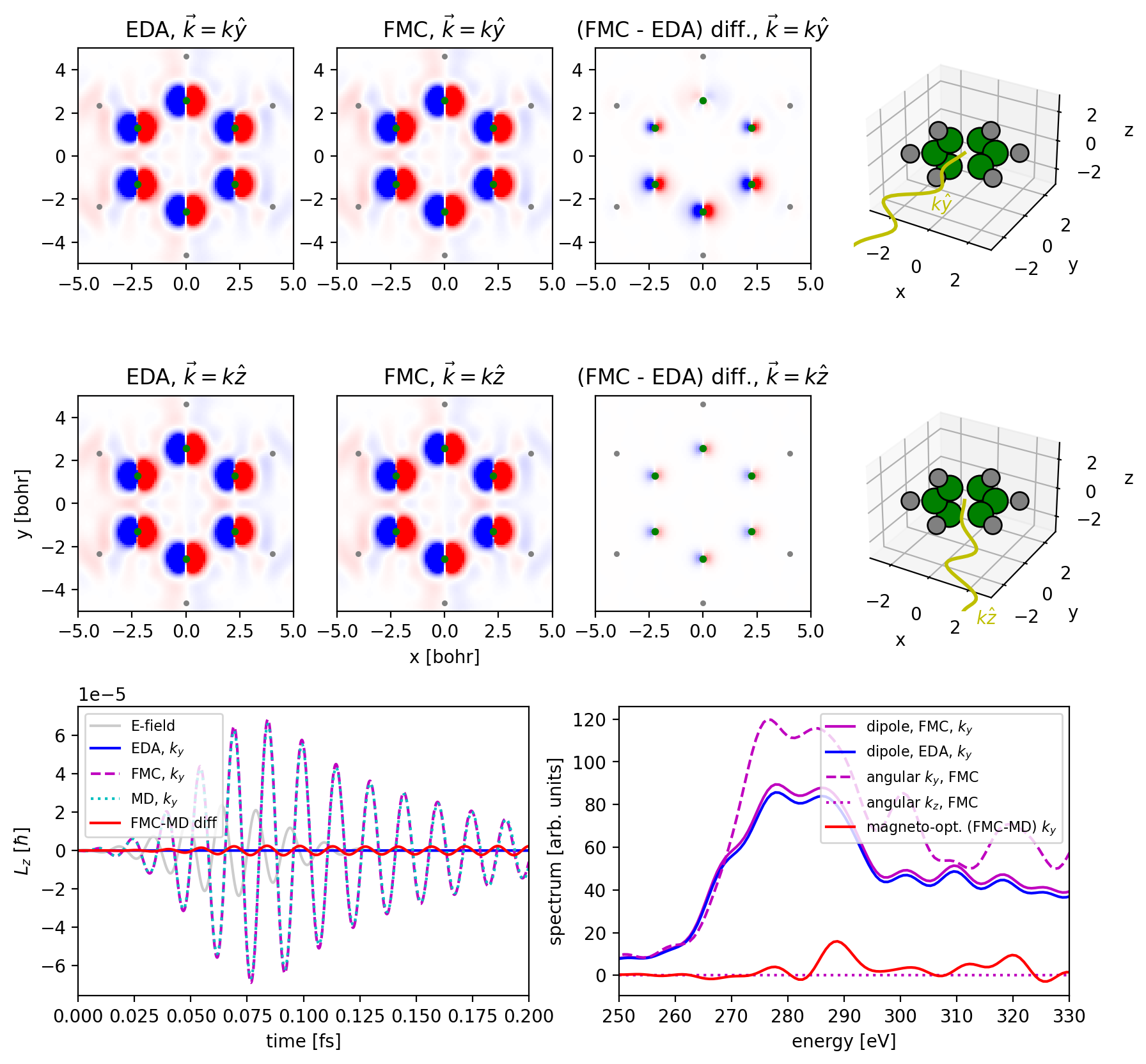}
    \caption{Laser-driven benzene at 270 eV in full minimal coupling (FMC) vs. electric dipole (ED) and mangetic dipole (MD): magneto-optical effects dependent on the incidence angle. The top panel shows snapshots of the electronic density in the $xy$ plane for thee case of the XUV pulse propagating along the $y$ direction. From the difference of the densities (FMC-ED), it can be seen that the atomic dipoles show a delay along the direction of propagation, which is a radiation pressure feature. This effect is not present when the angle of incidence is perpendicular to the molecule. A consequence of the retarded response along the propagation direction is a generation of angular momentum (bottom left). Subtracting the angular momentum induced by the magnetic dipole, the $L_z$ difference (red line) can be attributed to the magneto-optical effects. The spectrum of this cross response is ploted in the bottom-right panel, along with the (dipole) absorption spectra in FMC and ED, together with the angular spectra at different indicence angles, for comparison. It can be seen that the cross response tensor is mostly sensitive to one of the excitation peaks.}
    \label{fig:benzene_angular}
\end{figure}

The angular momentum expectation value is calculated as $\mathbf{L}(t) = \sum_{j=1}^{N_{\rm occ}} \int \mathrm{d}\mathbf{r} \psi_j^\dagger \hat{\mathbf{L}} \psi_j $. As shown in Fig. \ref{fig:benzene_angular}, angular momentum in the $z$ direction is driven when the full minimal coupling approach is used, but not in the dipolar coupling case. The induced magnetic moment $\mathbf{m} = \mathbf{L}$ (for a spin-unpolarized case) ignoring higher order terms can be written as \cite{Varsano2009}
\begin{align}
\mathbf{L}(\omega) = \boldsymbol{\chi}(\omega) \mathbf{B}(\omega) + \frac{i\omega}{c} \boldsymbol{\beta}(\omega) \mathbf{E}(\omega),
\label{eq:magnetization}
\end{align}
where $\mathbf{\chi}$ is the magnetic susceptibility tensor and $\mathbf{\beta}$ is the crossed response tensor. 

As the linear magnetic contribution is a trivial source of induced angular momentum, we subtract the magnetic dipole effect a simulation where the external field couples only through the magnetic dipole term (see eq. \eqref{eq:multipolarham}).
The difference between the full coupling and the magnetic dipolar coupling is the magnetooptical response (see Fig.~\ref{fig:benzene_angular}), which can be related to the time-domain magnetization as
\begin{align}
\mathbf{L}(t) - \mathbf{L}^{\mathrm{MD}}(t) -  = \frac{i\omega}{c} \int_{-\infty}^\infty \boldsymbol{\beta}(\omega)\mathbf{E}(\omega)  e^{-i\omega t} \mathrm{d}t,
\label{eq:magnetization2}
\end{align}
where $\mathbf{L}(t)$ and $\mathbf{L}^{\mathrm{MD}}(t)$ are $z$ components of the angular momenta calculated in full minimal coupling and at the magnetic dipole level, respectively.

For the calculation of circular dichroic spectra, only the diagonal elements of the crossed $\boldsymbol{\beta}$ tensor are considered to be relevant, since they are linked to the rotatory strength $ R(\omega) = \frac{3 \omega}{\pi c} \mathrm{Im}(\bar{\beta}(\omega))$. However, the off-diagonal components which vanish upon spherical averaging could play a role in novel beyond-dipole spectroscopies. In particular, here we compute the off-diagonal cross response as 
\begin{align}
\beta_{xz} (\omega) = \frac{1}{2\omega E_x(\omega)} \int_0^\infty (L_z(t) - L_z^{\mathrm{MD}}(t)) e^{i\omega t}\mathrm{d}t.
\label{eq:crosstensor}
\end{align}
The magneto-optical resonances due to the cross response can be visualized in the mangetooptical spectrum, which is compared with the dipolar (absorption) spectra (calculated as $\omega \Im(\langle x\rangle (\omega)/E_x(\omega))$) and with the angular spectra (defined as $\omega \Im (L_z (\omega) / E_x(\omega))$)  in the bottom-right panel of Fig. \ref{fig:benzene_angular}. While the dipole spectrum shows very similar peaks both in dipolar and FMC levels, with two major peaks around 278 and 290 eV, the magnetooptical spectrum shows only a clear resonance around 290 eV and other minor peaks. It is possible to see that the angular momentum spectrum calculated directly from the total $L_z$ does not show this selectivity of for the 290 nm excitation, allowing us to discriminate between magnetic dipole and higher-order (e.g. electric quadrupole) excitations. The reason for the selectivity of the magneto-optical spectrum on the second excited state needs still to be explored in depth, and is beyond the scope of this publication. In summary, the effects that arise from the higher-order coupling when using non-chiral light could be important to describe the origin of chirality in non-chiral oriented samples, especially in conditions where the dipole approximation is not valid (e.g. small wavelengths).

\section{Real-time radiation-reaction of plasmonic dimers}

Plasmonic nanostructures and nanocavities have proven to be useful platforms for driving non-equilibrium phenomena, including strong coupling \cite{Chikkaraddy2016}, ultrafast transport \cite{Ludwig2020}, nonlinear \cite{Rasmussen2023} and quadrupolar effects \cite{Sanders2018,Rivera2016}. The highly localized plasmonic fields at the apex of atomically-sharp tips are being used to achieve more and more resolved spectro-microscopy techniques, including TERS images \cite{Liu2019,Liu2023}. Precise knowledge of the spatial shape \cite{Barbry2015,Albar2023} and waveform \cite{Peller2021} of the plasmonic near fields is important to understand the mechanisms and push the envelope in the atomic-scale resolution techniques \cite{Siday2024}.

In this sense, radiation dynamics and radiative decay are often ignored aspects, which can lead to considerable frequency shifts when treated \cite{Downing2017,Downing2018} and could be used to tailor the modal structure of the radiative rate with respect to the excitation rates \cite{Bedingfield2022}. The mentioned frequency shift can be attributed to a nanoplasmonic analog of the atomic Lamb shift \cite{Downing2017}. Such a shift has also been demonstrated to arise in ab initio simulations with radiation reaction included at the mean-field level \cite{Schafer2023}. However, radiation damping and the associated Lamb-like shifts are expected to arise for large cluster sizes \cite{kreibig2013optical}.

\begin{figure}
    \centering
    \includegraphics[scale=0.5]{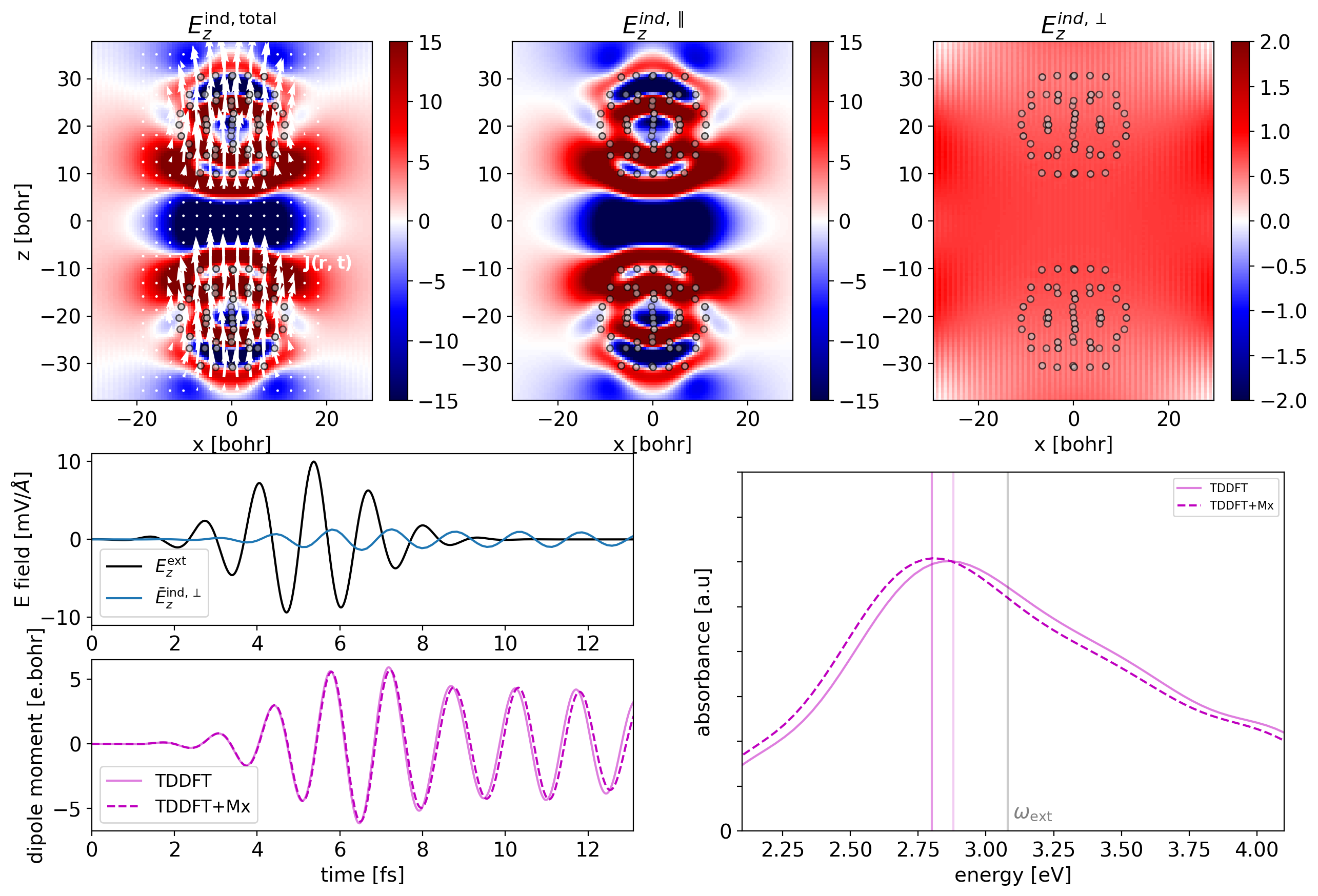}
    \caption{Transverse-field-effects on the plasmon-driven dynamics of a dimer of sodium clusters: total, longitudinal and transverse electric field spatial distribution at $t=7.5$ fs (top), transverse field (external and induced) dynamics and dipole moment (bottom-left), and spectra which evidence the renormalized plasmon frequency (bottom-right). }
    \label{fig:dimer}
\end{figure}

Here we demonstrate that even in free space, there is a potentially measurable renormalization of the dimer plasmonic mode in clusters of $\approx$ 1 nm diameter.  We have modeled a plasmonic dimer using two icosahedral sodium clusters of 55 atoms each, oriented in a face-to-face configuration and with an interparticle distance of 0.7 nm. We use a simulation box of 38$\times$38$\times$83 bohr$^3$ with a spacing of 0.28 bohr for the matter system, while for the Maxwell system the box is of size 113$\times$113$\times$159 bohr$^3$ and the spacing is 0.56 bohr (see Fig.~\ref{fig:dimer}, top left panel). The Maxwell box includes a PML boundary region of width 28 bohr in all directions. We propagate the dynamics for 13 fs using the exponential midpoint method, using timesteps of 0.5 as and 0.02 as for the matter and Maxwell systems, respectively,  under an external laser coupled in electric dipole approximation, tuned to the plasmonic resonance of the 55-atom cluster (3.08 eV). 

Fig. \ref{fig:dimer} shows a snapshot of the induced currents of the dimer at $t=7.5$ fs, along with the total induced electric field and its decomposition in longitudinal and transverse components. The transverse component is much smaller than the longitudinal one (which shows near-field enhancement with respect to the external field), but its amplitude reaches approximately 10\% of the external field. The transverse component is coupled back to obtain the renormalized radiative dynamics. The time-dependent dipole moment and the absorption spectra for the two simulations (with and without radiation reaction effects) are compared, where a frequency shift of the dipole plasmon mode of ca.~-70 meV is evident when the back-coupling is included, together with a broad absorption enhancement for energies below the absorption maximum.

\begin{figure}
    \centering
    \includegraphics[scale=0.4]{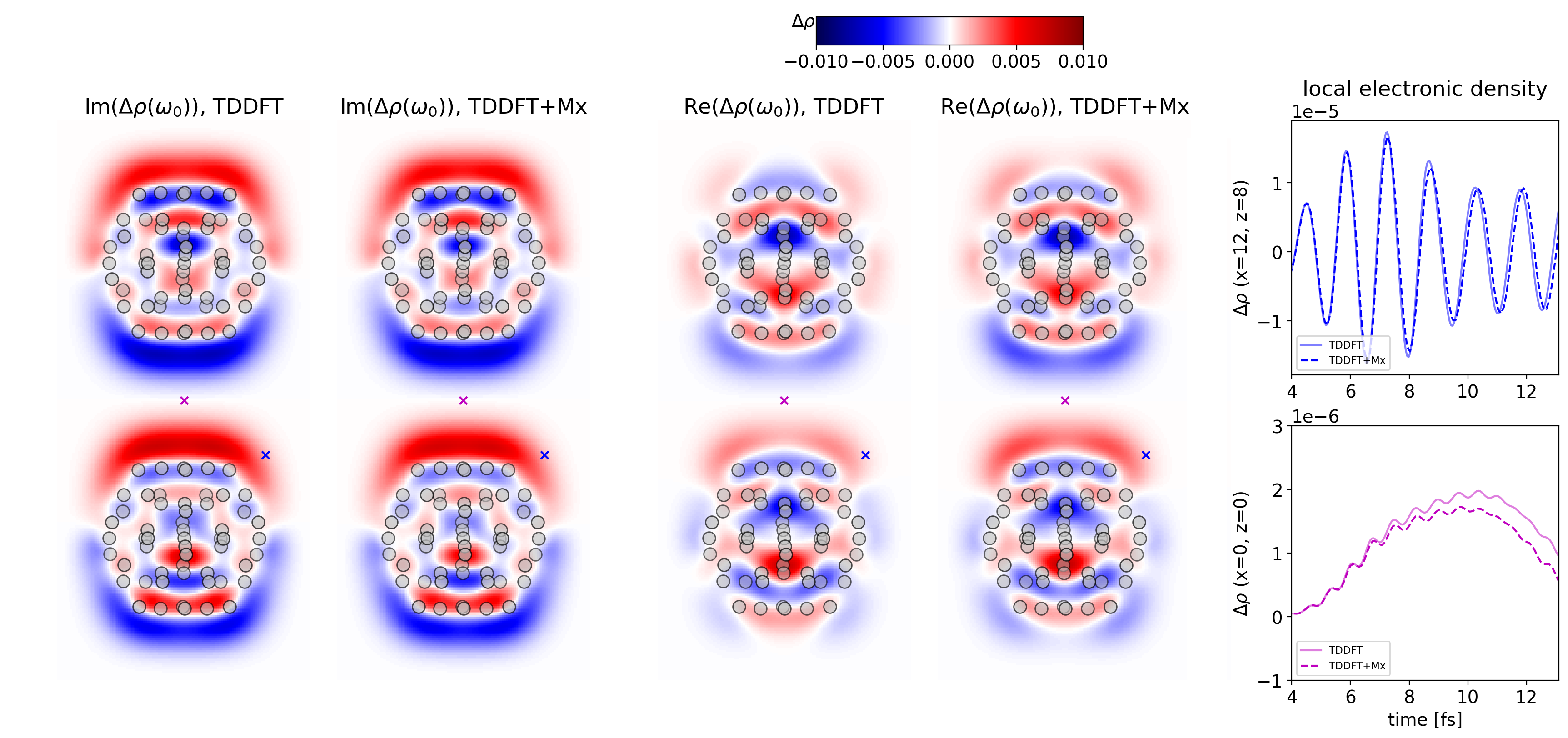}
    \caption{Plasmon-driven dynamics of a dimer of sodium clusters: Fourier-transformed local electronic density imaginary parts (left) and real parts (middle). Time-traces od the electronic density at selected points (right), evidencing the space-dependence of the shift in amplitude and phase.}
    \label{fig:dimer2}
\end{figure}

\begin{figure}
    \centering
    \includegraphics[scale=0.4]{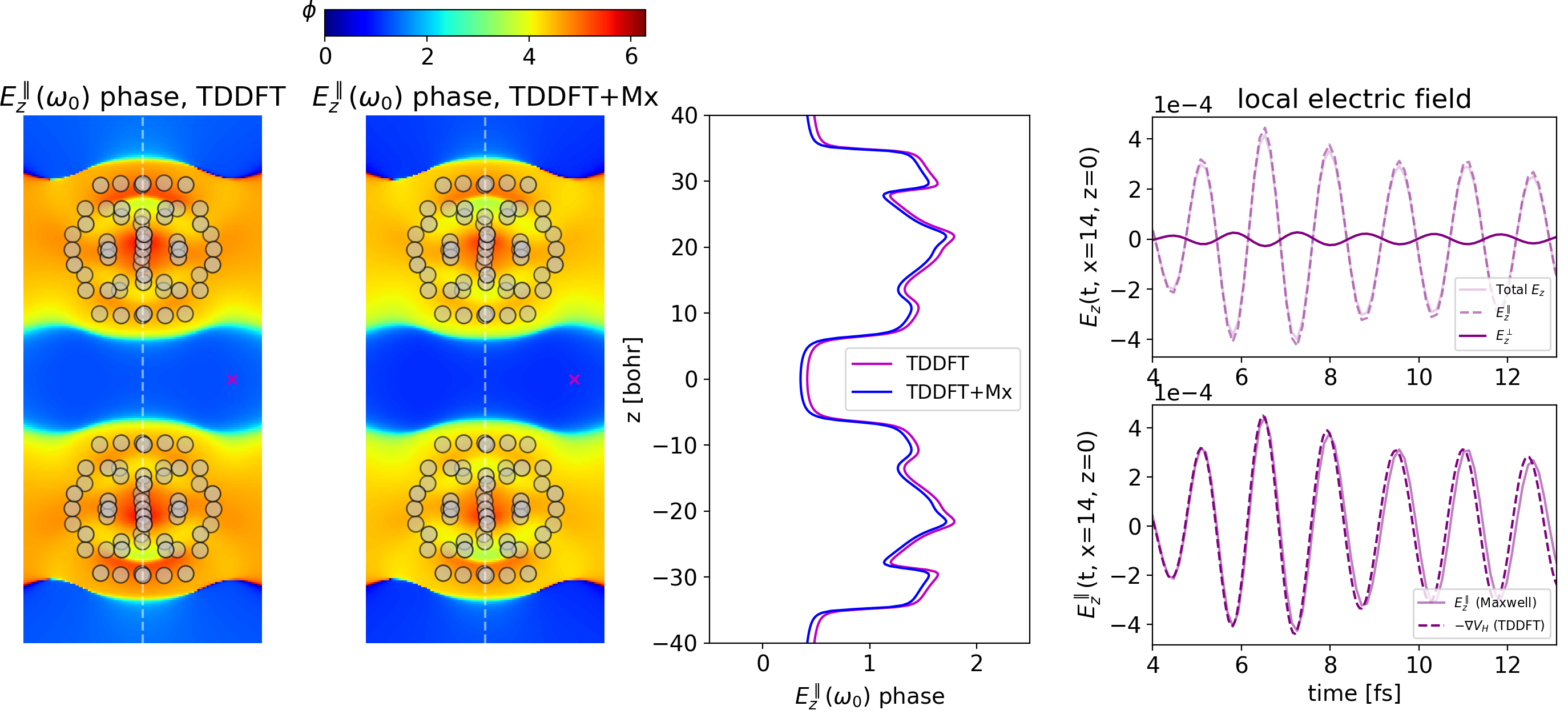}
    \caption{Plasmon-driven dynamics of a dimer of sodium clusters: spatial distribution phase of the Fourier-transformed longitundinal electric (near) field (slice in the $xz$ plane (left) and slice along the $z$ axis (middle). Components of the total electric field at a selected point (right-top), and time-trace of the near field with and without Maxwell coupling (right-bottom).}
    \label{fig:dimer3}
\end{figure}

The effects in real-space observables can be understood by comparing the local electronic density and near field components including the renormalization, depicted in Figs.~\ref{fig:dimer2} and \ref{fig:dimer3}. In the left panels the Fourier-transformed real-space electronic density in the $xy$ plane at $\omega_0 = 2.87$ eV is shown, along with the time trace of the density at selected points. The imaginary part of the density, which indicates the charge distribution of the driven plasmonic mode and is connected to light absorption \cite{Marchesin2016}, is not considerably affected by the radiation reaction. However, the real part does show non-trivial effects, evidencing a different scattering response. As seen in the time trace at specific points, in addition to the relative phase close to the surface of one of the clusters (dynamics at $x=12$ bohr, $z=8$ bohr), there are spill-out effects in the gap region, which could have implications for ultrafast electron transport in plasmonic gaps \cite{Ludwig2020}.

The near field calculated from the gradient of the Hartree potential is shown to be affected by the radiation reaction in Fig.~\ref{fig:dimer3}. The phase of the near field at the plasmonic mode frequency is shifted approximately $0.06\pi$ rad at the hot spot. As this quantity is affected by the plasmonic gap geometry and dielectric properties of the nanostructures, this phenomenon suggests radiative dynamics could play a role in experiments that depend on the quantum dynamics of plasmonic nanojunctions, e.g. in light-driven STM \cite{Peller2021} or TERS \cite{Liu2023}.

\section{Concluding remarks}

In summary, in this work we have proven that there are significant effects beyond dipole that play a role in the dynamics of quantum systems in electromagnetic environments. We have done so by implementing (to the best of our knowledge for the first time) a self-consistent Maxwell-TDDFT method in full minimal coupling with the total (external and induced) spatio-temporal electromagnetic fields. Generally speaking, the beyond-dipole response in mean-field QED will be relevant whenever (a) the transverse induced field is large compared to the external field, and (b) either the external or the induced vector potential has spatial inhomogeneities in the length scale of the system, which causes magnetic, or higher-order electric effects. Relevant examples of such conditions, which have not been covered in this article but will be studied in future works, are the excitation of quantum systems with orbital angular momentum beams (also called "twisted light"), strong field phenomena such as photoionization or high harmonic generation, x-ray absorption spectra using structured light, and plasmon-enhanced scattering spectra such as TERS. Overall, this new framework constitutes a powerful tool to assess the beyond-dipole effects as new knobs to control non-equilibrium phenomena by tailoring light-matter interactions in optical cavities.

\section*{Acknowledgements}
The authours would like to thank the Octopus developers team, especially Nicolas Tancogne-Dejean, Micael Oliveira, Martin Lüders and Alex Buccheri for contributing to code development and optimization, and Hans Fangohr, Victor Horvath and Andrea Petrocchi for their support on the use of the Postopus tool and their work on reproducibility. We also would like to thank Michael Ruggenthaler and Christian Schäfer for discussions. This work was supported by the European Research Council (ERC-2015-AdG694097), the Cluster of Excellence ‘Advanced Imaging of Matter' (AIM), Grupos Consolidados (IT1453-22) and Deutsche Forschungsgemeinschaft (DFG) – SFB-925 – project 170620586. The Flatiron Institute is a division of the Simons Foundation. We acknowledge support from the Max Planck-New York City Center for Non-Equilibrium Quantum Phenomena. F.P.B. acknowledges financial support from the European Union’s Horizon 2020 research and innovation program under the Marie Sklodowska-Curie Grant Agreement no. 895747 (NanoLightQD). E.I.A. acknowledges support from International Max Planck Research School.
The authors also would like to acknowledge the computational support provided by Max Planck Computing and Data Facility.

\bibliographystyle{unsrt}
\bibliography{biblio}

\end{document}


\title{Supporting Information for \\ 
``Full minimal coupling Maxwell-TDDFT: an ab initio framework for light-matter phenomena beyond the dipole approximation''
}

\author{Franco P. Bonafé\footnote[1]{Contact author: franco.bonafe@mpsd.mpg.de}}
\affiliation{Max Planck Institute for the Structure and Dynamics of Matter, Center for Free Electron Laser Science, Luruper Chaussee 149, 22761 Hamburg, Germany}
\author{Esra Ilke Albar}
\affiliation{Max Planck Institute for the Structure and Dynamics of Matter, Center for Free Electron Laser Science, Luruper Chaussee 149, 22761 Hamburg, Germany}
\author{Sebastian T. Ohlmann}
\affiliation{Max Planck Computing and Data Facility, Gießenbachstr. 2, 85748 Garching, Germany}
\author{Valeriia P. Kosheleva}
\affiliation{Max Planck Institute for the Structure and Dynamics of Matter, Center for Free Electron Laser Science, Luruper Chaussee 149, 22761 Hamburg, Germany}
\author{Carlos M. Bustamante}
\affiliation{Max Planck Institute for the Structure and Dynamics of Matter, Center for Free Electron Laser Science, Luruper Chaussee 149, 22761 Hamburg, Germany}
\author{Francesco Troisi}
\affiliation{Max Planck Institute for the Structure and Dynamics of Matter, Center for Free Electron Laser Science, Luruper Chaussee 149, 22761 Hamburg, Germany}
\author{Angel Rubio}
\affiliation{Max Planck Institute for the Structure and Dynamics of Matter, Center for Free Electron Laser Science, Luruper Chaussee 149, 22761 Hamburg, Germany}
\affiliation{Center for Computational Quantum Physics (CCQ), The Flatiron Institute, 162 Fifth Avenue, New York, New York 10010, USA}
\author{Heiko Appel\footnote[2]{Contact author: heiko.appel@mpsd.mpg.de}}
\affiliation{Max Planck Institute for the Structure and Dynamics of Matter, Center for Free Electron Laser Science, Luruper Chaussee 149, 22761 Hamburg, Germany}

\maketitle

\section{Simulation parameters for the harmonic oscillator}

For the electronic system a cubic box of size $20\times 20 \times 20$ bohr$^3$ and spacing of 0.15 bohr was used. Due to the harmonic potential, the wavepacket oscillations have a period $T=2\pi$ a.u. of time. Simultaneously, the Maxwell propagation was run, with the paramagnetic current from the electrons as the source term (as the diamagnetic and magnetization currents are zero). The Maxwell field is solved in a box of size $40 \times 40 \times 40$ bohr$^3$ using a spacing of 0.3 bohr. The absorbing PML region has a width of 9 bohr on all faces of the cube.

\section{Vector potential from the Riemann-Silberstein states}

The magnetic vector potential is calculated from the curl of the magnetic field $\mathbf{B}$, as shown in the main text. The implementation requires the use of a Poisson solver and the calculation of the numerical derivatives of the $\mathbf{B}$ field. As the derivatives will give approximate values close to the boundaries of the box, the vector potential could have a large error for situations in which the field is finite at the box boundaries (e.g. plane waves). We assess the accuracy of the vector potential in two ways: by re-computing the $\mathbf{B}$ field from its curl; and by comparing its time-derivative to yield the transverse electric field, as $\mathbf{E} = -\partial_t \mathbf{A}$ (in SI units). Here we show an example of the latter comparison, for a plane wave of frequency 270 eV propagating along the $y$ direction and polarized in the $x$ direction. The wave is fed in the box by means of plane wave conditions (Dirichlet boundary conditions), and the box is of 80x320x80 bohr$^3$. The results show that farther away from the boundaries, the error decreases, being smaller at 120 bohr with respect the origin of the box (280 bohr with respect to the incoming plane waves boundary). As the boundaries will induce numerical errors, the deviations are expected, and it is meant to be only a test. For all the calculations presented in this work, only the induced fields are propagated by means of the Riemann-Silberstein equations, therefore they decay to zero towards the boundaries, due to the use of absorbing boundaries (perfectly matches layer). Hence, these conditions ensure a good accuracy of the vector potential for the induced fields.

\begin{figure}
    \centering
    \includegraphics[scale=0.4]{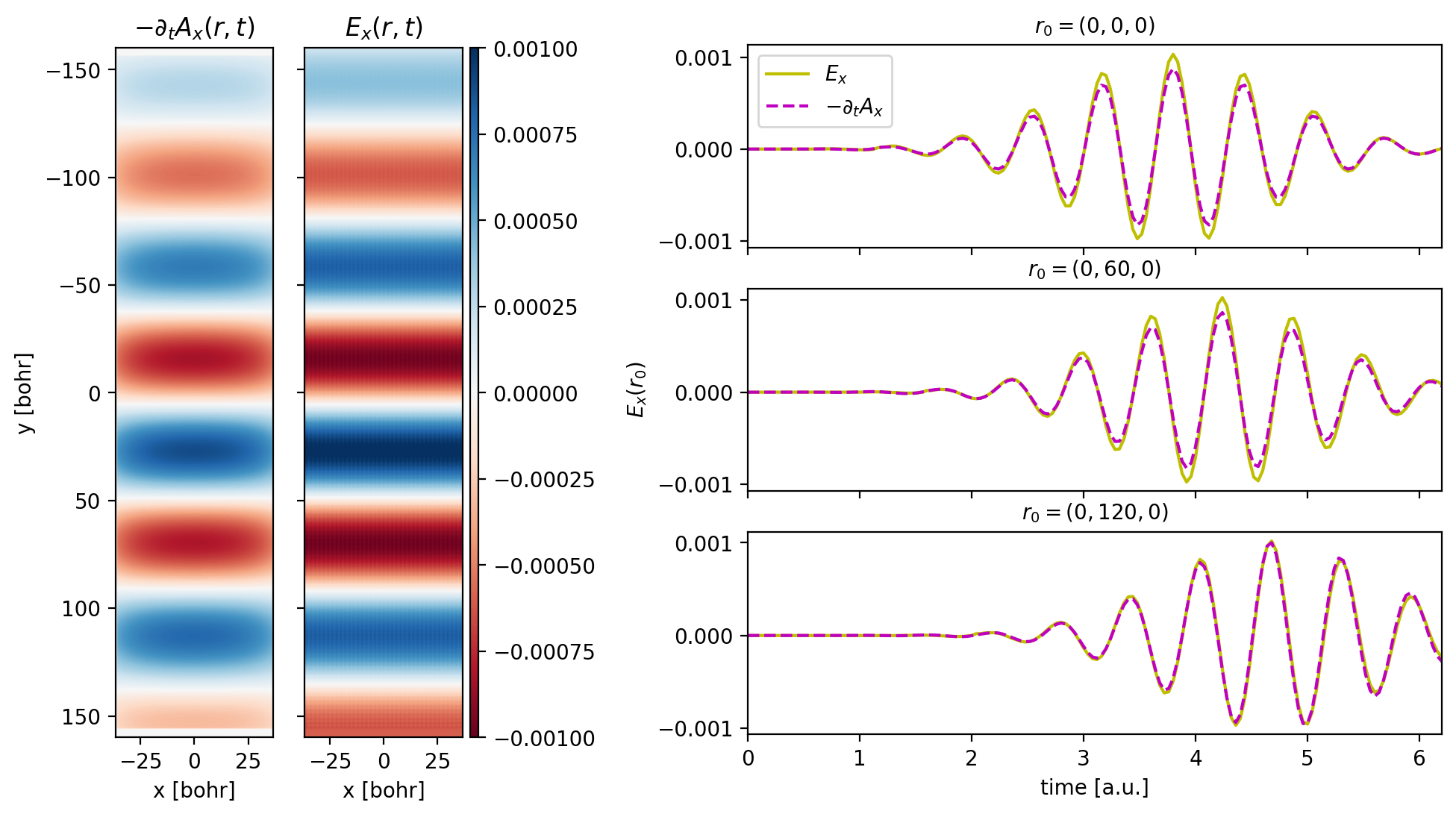}
    \caption{Comparison of the propagated electric field with the time derivative of the vector potential, to assess the accuracy of the magnetic vector potential calculation.}
    \label{fig:helm1}
\end{figure}

\section{Helmholtz decomposition}

\subsection{Methods and approximations}
The Helmholtz decomposition theorem states that any sufficiently smooth vector field can be expressed as the sum of an irrotational (curl-free) and a solenoidal (divergence-free) vector field. 
This means that in general, if one has a generic field $\mathbf{F}\left(\mathbf{r}\right) \in \mathcal{C}^1\left(\mathbb{R}^n\right)$, then
\begin{align}
    \label{eq:helm_general}
    & \mathbf{F}\left(\mathbf{r}\right) = \mathbf{F}_{\mathrm{long}}\left(\mathbf{r}\right) + \mathbf{F}_{\mathrm{trans}}\left(\mathbf{r}\right) = -\nabla \phi\left(\mathbf{r}\right) + \nabla \times \mathbf{A}\left(\mathbf{r}\right) \\
    \label{eq:helm_long_trans}
    & \nabla \times \mathbf{F}_{\mathrm{long}}\left(\mathbf{r}\right) = 0,  
    \nabla \cdot \mathbf{F}_{\mathrm{trans}}\left(\mathbf{r}\right) = 0
\end{align}
\noindent where $\phi\left(\mathbf{r}\right)$ is a scalar function (scalar potential) and $\mathbf{A}\left(\mathbf{r}\right)$ is a vector field (vector potential).
In equation \ref{eq:helm_general} we exploited the conditions on the transverse and longitudinal fields (\ref{eq:helm_long_trans}) and some vectorial identities to write the final expression.
To give a more physical interpretation to the equations above, let us call $\mathbf{F}_{trans}\left(\mathbf{r}\right) = \mathbf{B}\left(\mathbf{r}\right)$ (magnetic field).

\noindent The theorem states that it is possible, given the total field $\mathbf{F}\left(\mathbf{r}\right)$ to compute its transverse and longitudinal components as:

\begin{equation}
    \label{eq:scalar_potential}
    \phi\left(\mathbf{r}\right) = \frac{1}{4\pi} \left( \int \frac{\nabla' \cdot \mathbf{F}\left(\mathbf{r}'\right)}{|\mathbf{r} - \mathbf{r}'|} \,d\mathbf{r}' - \oint \mathbf{\hat{n}}' \cdot \frac{\mathbf{F}\left(\mathbf{r}'\right)}{|\mathbf{r} - \mathbf{r}'|} \,dS' \right)
\end{equation}

\begin{equation}
    \label{eq:vector_potential}
    \mathbf{A}\left(\mathbf{r}\right) = \frac{1}{4\pi} \left( \int \frac{\nabla' \times \mathbf{F}\left(\mathbf{r}'\right)}{|\mathbf{r} - \mathbf{r}'|} \,d\mathbf{r}' - \oint \mathbf{\hat{n}}' \times \frac{\mathbf{F}\left(\mathbf{r}'\right)}{|\mathbf{r} - \mathbf{r}'|} \,dS'  \right)
\end{equation}

\noindent In both equations two terms appear, a volume integral and a surface integral. The former is computed by first applying the appropriate differential operator to $\mathbf{F}\left(\mathbf{r}\right)$, and then by solving the Poisson equation.
Note that in the case of the vector potential $\mathbf{A}\left(\mathbf{r}\right)$ we solve three Poisson equations, one for each direction. We tested our implementation using the Conjugate Gradient Poisson solver, however the user may choose any non-periodic algorithm (i.e. FFT should not be used).
Conversely, the surface integral can be neglected if the integration domain is large enough.
This approximation works well when the simulation box is sufficiently large, however in general neglecting the surface integral can lead to a potential that diverges greatly from the correct result.

\noindent The equations above do not make any assumption on the gauge of the field. Explicitly taking into account the gauge used in this work (the Coulomb gauge $\mathbf{\nabla} \cdot \mathbf{A} = 0$) allows us to simplify the expressions \cite{Stewart2003}. Equation \ref{eq:vector_potential} can be rewritten as:
\begin{equation}
    \label{eq:vector_potential_from_b}
    \mathbf{A}\left(\mathbf{r}\right) = \nabla \times \frac{1}{4\pi} \int \frac{\mathbf{B}\left(\mathbf{r}'\right)}{|\mathbf{r} - \mathbf{r}'|} \,d\mathbf{r}' 
\end{equation}
Due to numerical reasons, this expression is more stable to evaluate than equation \ref{eq:vector_potential}.
This is because Octopus uses finite differences to compute derivatives. Thus, the algorithm is well-defined in the inner regions of the simulation box, whereas at the boundaries one needs to include extra points (that are not physical). The artifact is that Octopus assumes any function to be zero in those points (unless the value is manually set).
This results in a jump in the value of the fields at the boundaries, thus generating spikes in the derivatives and leading to numerical instabilities both when computing the Poisson equation for $\nabla' \times \mathbf{F}\left(\mathbf{r}'\right)$ and when using $\mathbf{A}\left(\mathbf{r}\right)$ in the following time step(s).
In contrast, in equation \ref{eq:vector_potential_from_b}, the first operation to be computed is solving the Poisson equation for the magnetic field $\mathbf{B}\left(\mathbf{r}\right)$, which does not introduce any spike at the boundaries even if the field is non-zero in those points.
Therefore, one can still assume that the result of the Poisson equation is numerically correct. Moreover, before applying the curl we subtract from the field its mean value at the boundaries. Since this is a constant, it does not affect the curl operation and makes the value of the field at the boundaries closer to zero, thus generating fewer spikes after the curl.
As a result, all figures in the following discussion are generated using equation \ref{eq:vector_potential_from_b}. It is important to remember that despite equation \ref{eq:vector_potential_from_b} performing better than equation \ref{eq:vector_potential}, it cannot exactly compute the vector potential in all scenarios, thus in the following we show some examples.

\subsection{Accuracy of the Helmholtz decomposition of the field generated by a current source}

\begin{figure}
    \centering
    \includegraphics[scale=0.4]{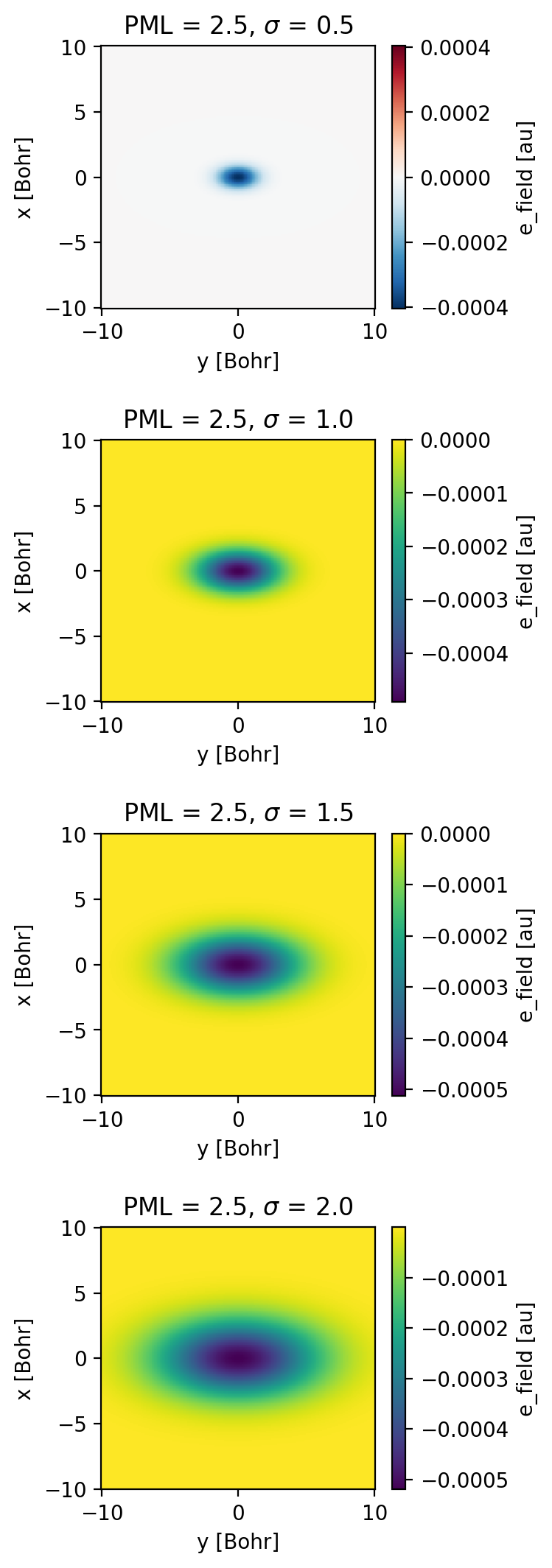}
    \caption{Electric field $E_z$ profile produced by the current densities with different width $\sigma$}
    \label{fig:ez_as_func_of_sigma}
\end{figure}

\begin{figure}
    \centering
    \includegraphics[scale=0.4]{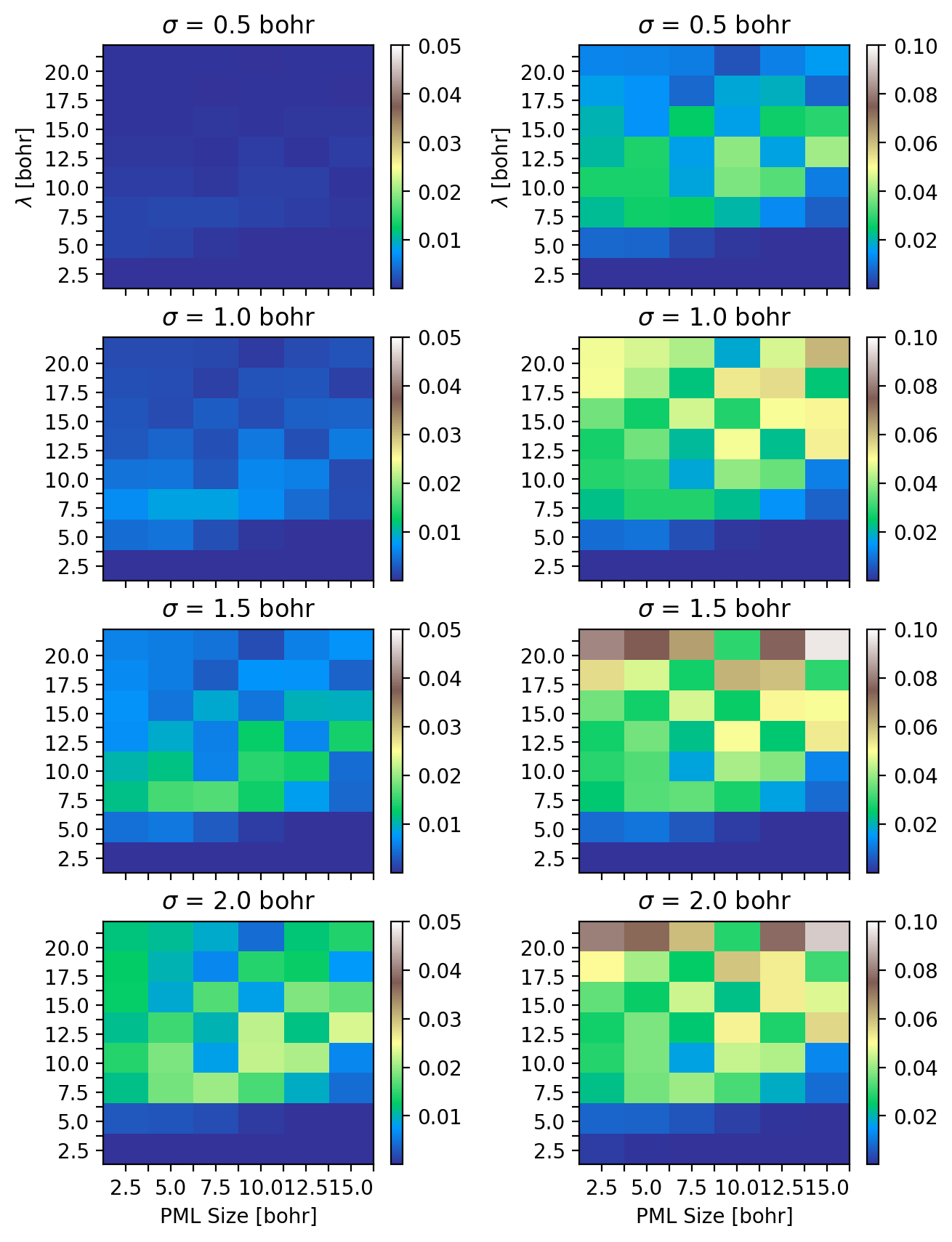}
    \caption{Absolute (left) and relative (right) error, defined as $Err = \left\lVert \int_{\mathcal{V}} \vec{E} - \vec{E}^\perp - \vec{E}^\parallel \right\rVert$ for different wavelength values $\lambda$, current density size $\sigma$ and PML width.}
    \label{fig:error_as_func_of_sigma}
\end{figure}

In Fig.~\ref{fig:error_as_func_of_sigma} one can see the absolute and relative error of the decomposed field at the final timestep as a function of the wavelength $\lambda$, the width of the PML region (or absorbing boundary region) and the Gaussian broadening $\sigma$.
One can immediately observe that the larger the value of $\sigma$, the bigger the error is. In particular, for $\sigma = 0.5, 1.0$ the plot shows that the error is normally less than 0.01 a.u., while for bigger values green and yellow spots appear. This can be explained by looking at Figure \ref{fig:ez_as_func_of_sigma}. For $\sigma = 0.5, 1.0$ the electric field is well contained in the simulation box, thus its value at the boundaries decays to zero, whereas for $\sigma = 1.5, 2.0$ the electric field distribution reaches the box boundaries.

\begin{figure}
    \centering
    \includegraphics[scale=0.4]{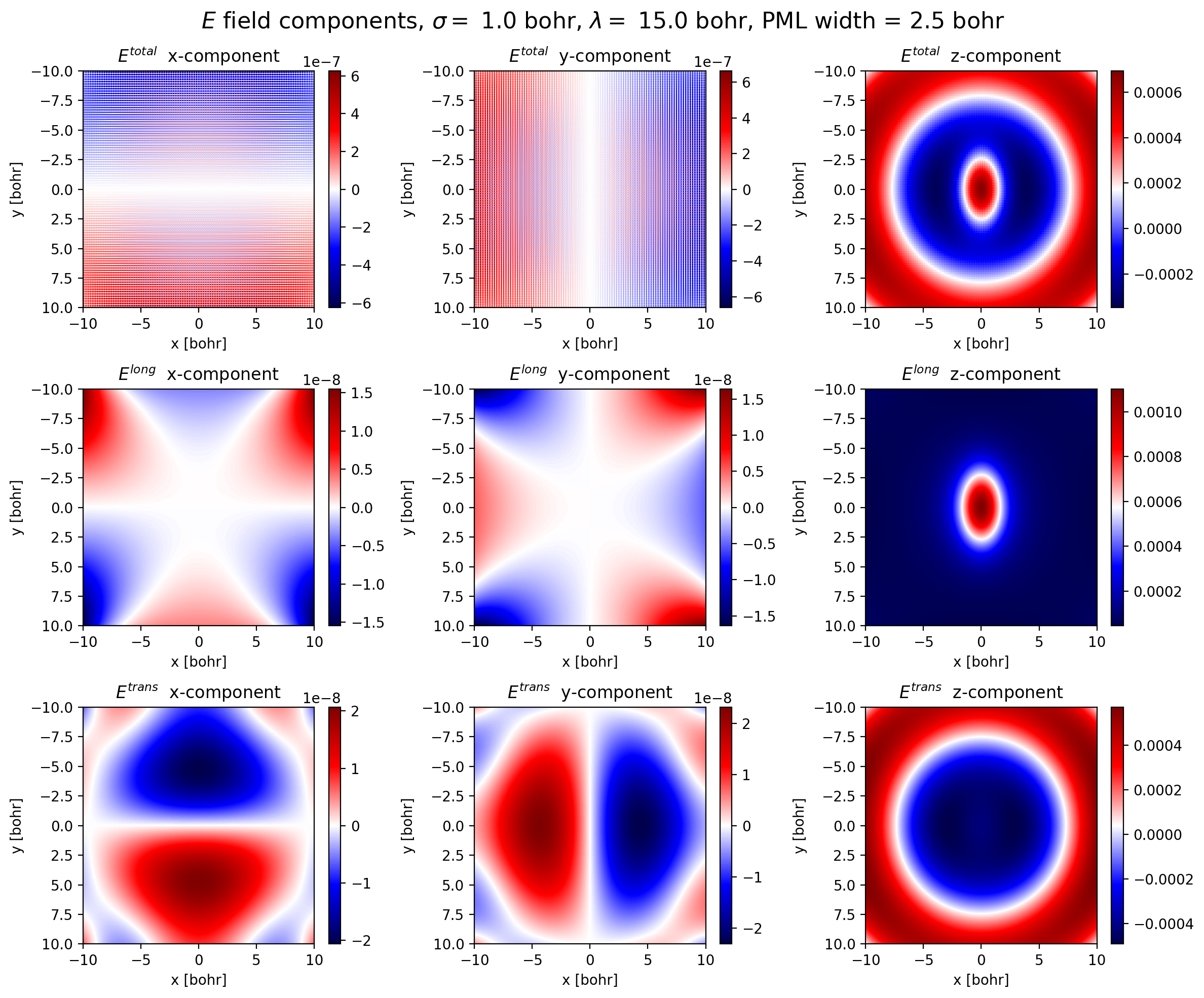}
    \caption{$x$, $y$ and $z$ components of the electric field and their calculated Helmholtz decomposition in longitudinal $\vec{E}^\parallel$ and transverse $\vec{E}^\perp$}
    \label{fig:helm_lam_15_pml_2.5}
\end{figure}

This can be well observed in Figure \ref{fig:helm_lam_15_pml_2.5}, where in the $x$ and $y$ components the decomposed fields are quite large at the boundaries.

\begin{figure}
    \centering
    \includegraphics[scale=0.4]{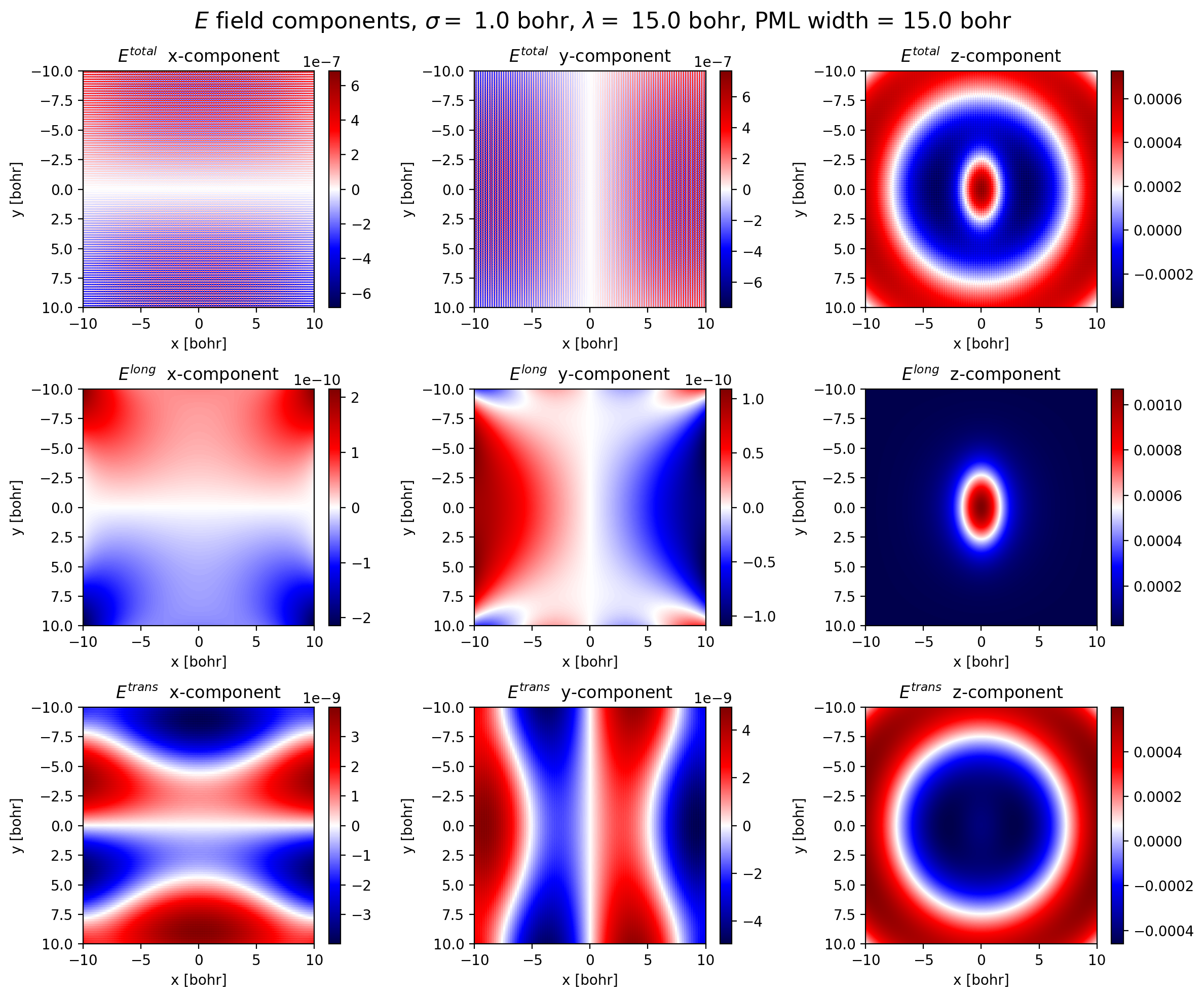}
    \caption{$x$, $y$ and $z$ components of the electric field and their calculated Helmholtz decomposition in longitudinal $\vec{E}^\parallel$ and transverse $\vec{E}^\perp$}
    \label{fig:helm_lam_15_pml_15}
\end{figure}

Another trend that can be observed is that the error in the decomposition tends to be smaller for larger widths of the PML region. There, the field is forced to decay to zero. Thus having a larger width implies that the field can go to zero more smoothly, especially at the beginning of the PML region, generating fewer spikes in the derivatives.
This can be seen by comparing the $x$ and $y$ components of the decomposed field in figures \ref{fig:helm_lam_15_pml_15} and \ref{fig:helm_lam_15_pml_2.5}. In the former, where the field has more space to decay, the decomposition (rows 2 and 3) can reproduce the horizontal or vertical bands in the exact field (row 1).
Despite increasing the PML region seems advantageous, one should remember that the calculation becomes more expensive (as the number of grid points increases). Thus, one should look for a good trade-off between accuracy and performance.

\begin{figure}
    \centering
    \includegraphics[scale=0.4]{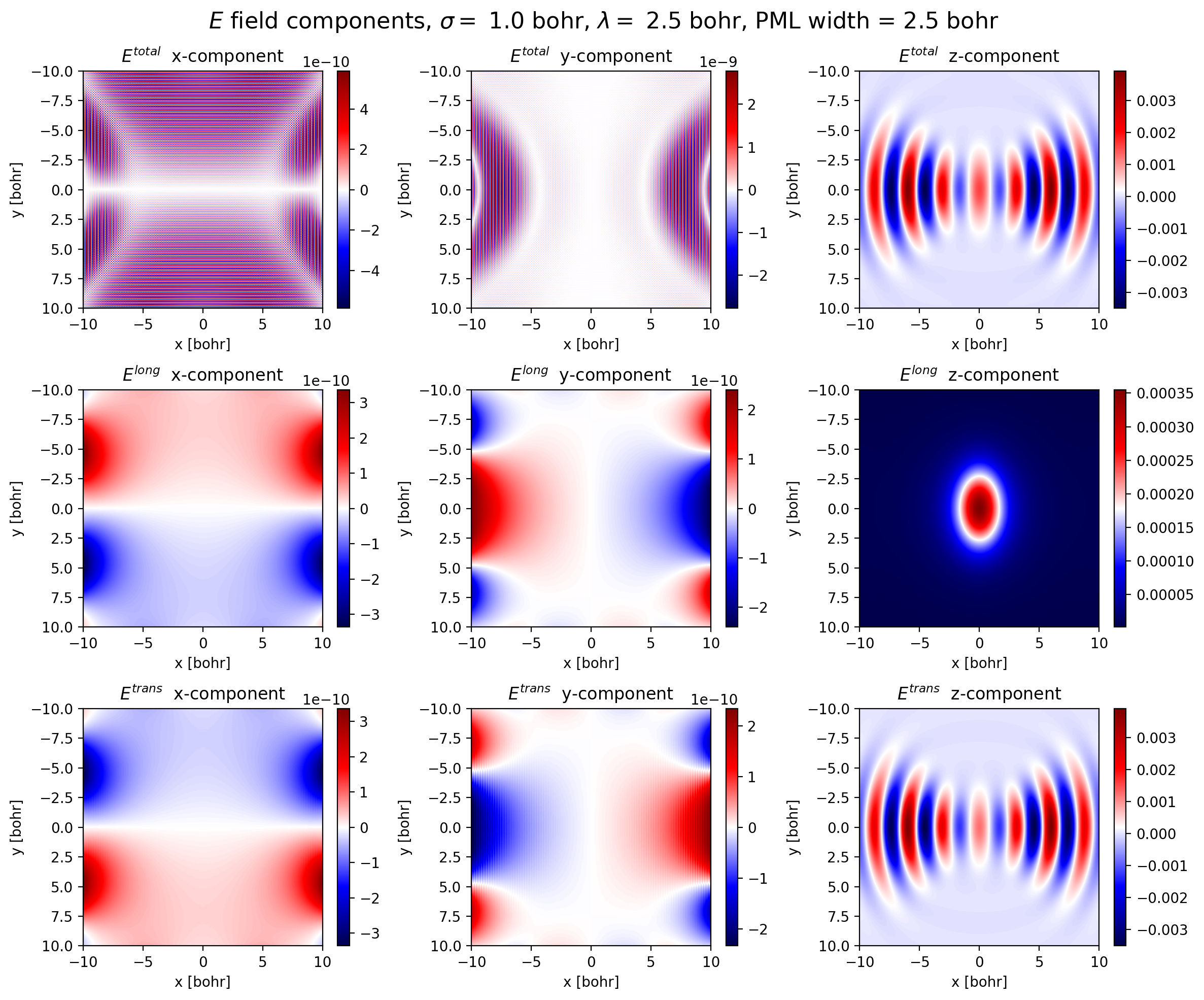}
    \caption{$x$, $y$ and $z$ components of the electric field and their calculated Helmholtz decomposition in longitudinal $\vec{E}^\parallel$ and transverse $\vec{E}^\perp$}
    \label{fig:helm_lam_2.5_pml_2.5}
\end{figure}

Finally, one can observe that the bigger the wavelength, the bigger the error. This can be understood if one imagines that in general having a bigger $\lambda$ implies that the Gaussian envelope is larger, thus the value of the field at the boundaries is larger.
Naturally, one can still have a good decomposition for a big $\lambda$ if it is a divisor of the simulation box (which explains the blue points in Figure \ref{fig:error_as_func_of_sigma} for $\lambda = 10.0$ bohr).
The effect of the wavelength can be seen by comparing figures \ref{fig:helm_lam_2.5_pml_2.5} and \ref{fig:helm_lam_15_pml_2.5}. In both, the PML region is quite small, however in the former the field is perfectly decomposed.

\begin{figure}
    \centering
    \includegraphics[scale=0.4]{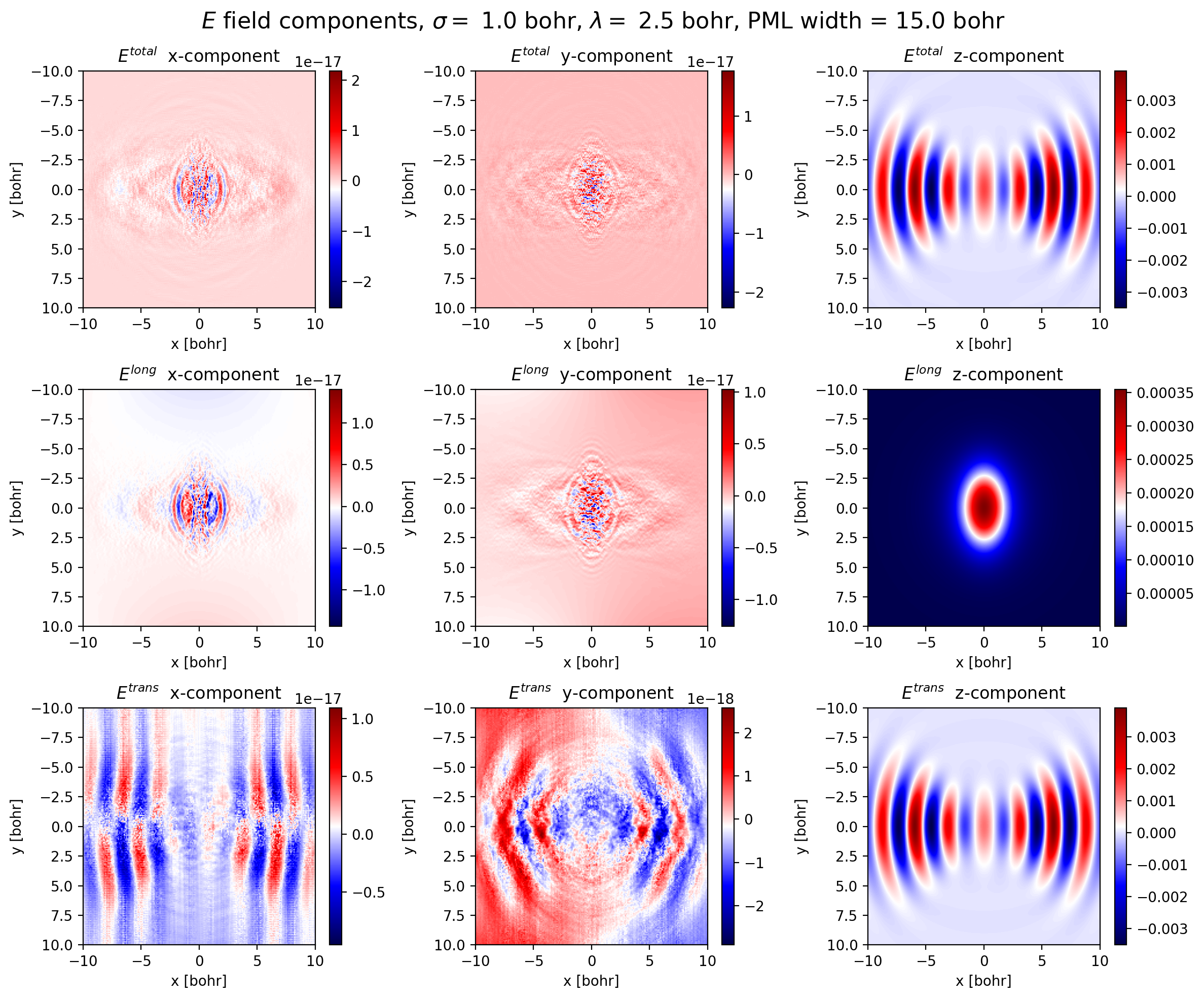}
    \caption{$x$, $y$ and $z$ components of the electric field and their calculated Helmholtz decomposition in longitudinal $\vec{E}^\parallel$ and transverse $\vec{E}^\perp$}
    \label{fig:helm_lam_2.5_pml_15}
\end{figure}

\bibliographystyle{unsrt}
\bibliography{biblio}